\documentclass[iop]{emulateapj}
\usepackage{amsmath,wasysym}
\usepackage{array,multirow,color,tablefootnote}
\usepackage[flushleft]{threeparttable}
\usepackage{float}
\usepackage{graphicx}
\usepackage{subfigure}
\usepackage{setspace}
\usepackage{epstopdf,textcomp}
\usepackage{natbib}

\newcommand{\ts}{\textsuperscript}
\newcommand{\bjdtdb}{\ensuremath{\rm {BJD_{TDB}}}}

\newcommand{\ms}{m s$^{-1}$}

\bibpunct[ ]{(}{)}{;}{}{}{,}
\shorttitle{WASP-19b and HAT-P-7b Phase Curves}
\shortauthors{Wong et al.}

\begin{document}
\title{3.6 and 4.5~$\mu$\MakeLowercase{m} \textit{Spitzer} Phase Curves of the \\ Highly Irradiated Hot Jupiters WASP-19\MakeLowercase{b} and HAT-P-7\MakeLowercase{b}}

\author{Ian Wong,\altaffilmark{1} Heather A. Knutson,\altaffilmark{1} Tiffany Kataria,\altaffilmark{2} Nikole K. Lewis,\altaffilmark{3} Adam Burrows,\altaffilmark{4} Jonathan J. Fortney,\altaffilmark{5} Joel Schwartz,\altaffilmark{6} Avi Shporer,\altaffilmark{7,1,17} Eric Agol,\altaffilmark{8} Nicolas B. Cowan,\altaffilmark{9,10} Drake Deming,\altaffilmark{11}  Jean-Michel D{\' e}sert,\altaffilmark{12}  Benjamin J. Fulton,\altaffilmark{13,18} Andrew W. Howard,\altaffilmark{13}  Jonathan Langton,\altaffilmark{14} Gregory Laughlin,\altaffilmark{5}  Adam P. Showman,\altaffilmark{15} and Kamen Todorov\altaffilmark{16}}
\affil{\textsuperscript{1}Division of Geological and Planetary Sciences, California Institute of Technology, Pasadena, CA 91125, USA; iwong@caltech.edu}
\affil{\textsuperscript{2}Astrophysics Group, School of Physics, University of Exeter, Stocker Road, Exeter EX4 4QL, UK}
\affil{\textsuperscript{3}Space Telescope Science Institute, Baltimore, MD 21218, USA}
\affil{\textsuperscript{4}Department of Astrophysical Sciences, Princeton University, Princeton, NJ 08544, USA}
\affil{\textsuperscript{5}Department of Astronomy and Astrophysics, University of California at Santa Cruz, Santa Cruz, CA 95604, USA}
\affil{\textsuperscript{6}Department of Physics \& Astronomy, Northwestern University, Evanston, IL 60208, USA}
\affil{\textsuperscript{7}Jet Propulsion Laboratory, California Institute of Technology, 4800 Oak Grove Drive, Pasadena, CA 91109, USA}
\affil{\textsuperscript{8}Department of Astronomy, University of Washington, Seattle, WA 98195, USA}
\affil{\textsuperscript{9}Department of Physics, McGill University, 3600 rue University, Montreal, QC, H3A 2T8, Canada}
\affil{\textsuperscript{10}Department Earth \& Planetary Sciences, McGill University, 3450 rue University, Montreal, QC, H3A 0E8, Canada}
\affil{\textsuperscript{11}Department of Astronomy, University of Maryland, College Park, MD 20742, USA}
\affil{\textsuperscript{12}Department of Astrophysical and Planetary Science, University of Colorado, Boulder, CO 80309, USA}
\affil{\textsuperscript{13}Institute for Astronomy, University of Hawaii, Honolulu, HI 96822,USA}
\affil{\textsuperscript{14}Department of Physics, Principia College, Elsah, IL 62028, USA}
\affil{\textsuperscript{15}Lunar and Planetary Laboratory, University of Arizona, Tucson, AZ 85721, USA}
\affil{\textsuperscript{16}Institute for Astronomy, ETH Z{\" u}rich, 8093 Z{\" u}rich, Switzerland}
\altaffiltext{17}{Sagan Fellow}
\altaffiltext{18}{NSF Graduate Research Fellow}

\begin{abstract}
We analyze full-orbit phase curve observations of the transiting hot Jupiters WASP-19b and HAT-P-7b  at 3.6 and 4.5~$\mu$m obtained using the Spitzer Space Telescope. For WASP-19b, we measure secondary eclipse depths of $0.485\%\pm 0.024\%$ and $0.584\%\pm 0.029\%$ at 3.6 and 4.5~$\mu$m, which are consistent with a single blackbody with effective temperature $2372 \pm 60$~K. The measured 3.6 and 4.5~$\mu$m secondary eclipse depths for HAT-P-7b are $0.156\%\pm 0.009\%$ and $0.190\%\pm 0.006\%$, which are well described by a single blackbody with effective temperature $2667\pm 57$~K. Comparing the phase curves to the predictions of one-dimensional and three-dimensional atmospheric models, we find that WASP-19b's dayside emission is consistent with a model atmosphere with no dayside thermal inversion and moderately efficient day--night circulation. We also detect an eastward-shifted hotspot, which suggests the presence of a superrotating equatorial jet. In contrast, HAT-P-7b's dayside emission suggests a dayside thermal inversion and relatively inefficient day--night circulation; no hotspot shift is detected. For both planets, these same models do not agree with the measured nightside emission. The discrepancies in the model-data comparisons for WASP-19b might be explained by high-altitude silicate clouds on the nightside and/or high atmospheric metallicity, while the very low 3.6~$\mu$m nightside planetary brightness for HAT-P-7b may be indicative of an enhanced global C/O ratio. We compute Bond albedos of $0.38\pm 0.06$ and 0 ($<0.08$ at $1\sigma$) for WASP-19b and HAT-P-7b, respectively. In the context of other planets with thermal phase curve measurements, we show that WASP-19b and HAT-P-7b fit the general trend of decreasing day--night heat recirculation with increasing irradiation.
\end{abstract}

\section{Introduction}\label{sec:intro}

Over the past decade, phase curve observations have proven to be an invaluable tool in the intensive atmospheric characterization of an increasingly diverse  body of exoplanets. By measuring the variation in the observed infrared flux of the system throughout an orbit, one obtains the planet's emission across a range of viewing geometries, from which a longitudinal temperature map of the planet's surface can be acquired \citep{cowanagol}. Basic properties of the atmosphere can be deduced by examining the shape of the phase curve: the amplitude of the phase curve modulation provides the day--night temperature contrast, while the measured phase offset of the brightness maximum relative to the time of secondary eclipse reveals longitudinal shifts in the emission maximum relative to the substellar point, which is related to the interplay between the atmospheric advective and radiative timescales \citep[e.g.,][]{showman2009,lewis2010,perez-becker}. Together, these characteristics describe the thermal budget of the planet's atmosphere and help constrain such properties as the Bond albedo and the recirculation efficiency \citep{schwartz}.

Multiband phase curve observations add a new vertical dimension to our understanding of a planet's atmosphere. The measured modulation in the planet's emission at each wavelength provides information about the atmospheric energetics and dynamics at a particular height within the atmosphere \citep[e.g.][]{knutson2012,lewis}. To date, only six planets have well-characterized phase curve detections at more than one wavelength: HD 189733b \citep{knutson2012}, WASP-12b \citep{cowan2012}, WASP-18b \citep{maxted}, HAT-P-2b \citep{lewis}, WASP-43b \citep{stevenson}, and WASP-14b \citep{wong2}.

A full interpretation of multiband phase curve observations requires the comparison of the measured fluxes with the results of theoretical models. The detailed morphology of the phase variation and the relative planetary brightness in various bandpasses (i.e., phase-resolved broadband emission spectra) are affected by many aspects of the atmospheric dynamics and composition. By matching the phase curve data with model-generated light curves and spectra, one can infer the presence or absence of a dayside temperature inversion, atmospheric metallicity, C/O ratio, wind speeds, as well as the presence of clouds and/or hazes \citep[e.g.,][]{fortney2006,showman2009,lewis2010,lewis,hengdemory,madhusudhan}.

While the complexity and explanatory capabilities of atmospheric models have increased over the past decade, reproducing multiband phase curve observations with theoretical light curves remains a challenge. In particular, while the dayside planetary emission is typically well-matched by current models, the observed nightside emission at various wavelengths often diverges significantly from the theoretical predictions \citep[e.g.][]{lewis,wong2}, underscoring the need for improved models in order to better understand the underlying physical and chemical processes that drive atmospheric circulation. Nevertheless, the growing body of exoplanet phase curves has begun to reveal several notable trends.  Comparisons of the day--night temperature contrasts for exoplanets with thermal phase curve observations \citep[e.g.,][]{cowanagol2011}, as well as hydrodynamical simulations \citep{perna2,perez-becker}, have shown that planets with higher levels of stellar irradiation have systematically poorer day--night heat recirculation. An additional trend is seen in the distribution of estimated albedos: using a simple thermal balance to solve for the Bond albedo of planets with thermal phase curve measurements, following the methods of \citet{schwartz}, we showed in \citet{wong2} that the albedo distribution is bimodal, with the massive hot Jupiters ($7-10~M_{\mathrm{Jup}}$) having albedos consistent with zero, and the other less-massive hot Jupiters having higher albedos.

To explore further these emerging trends, we have obtained phase curve observations of the hot Jupiters WASP-19b and HAT-P-7b in the 3.6 and 4.5~$\mu$m bands using the Spitzer Space Telescope. Photometric and radial velocity (RV) observations of WASP-19b indicate a mass of $M_{p}=1.114\pm0.036~ M_{\mathrm{Jup}}$ \citep{tregloanreed}. The planet orbits a G8-type host star with a period of 0.789~days \citep{tregloanreed}. Transmission spectroscopy studies of WASP-19b from the optical to the near-infrared reveal a strong detection of water vapor in the atmosphere as well as a weak or absent dayside temperature inversion \citep{bean,huitson}. Atmospheric retrieval using the transmission spectroscopy data rules out high C/O ratios \citep{benneke}. Analyses of the dayside thermal emission via occultation light curves corroborate the lack of a strong temperature inversion \citep{anderson,lendl,zhou}. HAT-P-7b has a mass of $M_{p}=1.776^{+0.077}_{-0.049}~ M_{\mathrm{Jup}}$ \citep{pal} and orbits an F8-type host star with a period of 2.20~days \citep{morris}. Previous analyses of a full-orbit optical \textit{Kepler} phase curve as well as secondary eclipse light curves in the optical and infrared suggest the presence of a strong dayside temperature inversion and inefficient day--night heat recirculation \citep{borucki,christiansen,spiegel}.

Both planets lie on nearly circular orbits and have similar levels of incident stellar radiation and gravitational acceleration, which are two key drivers of atmospheric circulation. A full list of relevant stellar, planetary, and orbital properties for our two target systems is provided in Table~\ref{tab:properties}. A way to quantify the intensity of the incident stellar irradiation is to calculate the irradiation temperature $T_{0}\equiv T_{*}\sqrt{R_{*}/a}$ at the substellar point, which is defined such that $F_{0}=\sigma T_{0}^{4}$ is the incident stellar flux, where $\sigma$ is the Stefan--Boltzmann constant. WASP-19b and HAT-P-7b have irradiation temperatures of 3000~K and 3210~K, respectively, making them among the most highly irradiated hot Jupiters known. 

The paper is organized as follows. The observations, data reduction techniques, and phase curve model are summarized in Section~\ref{sec:obs}. In Section~\ref{sec:analysis}, we present phase curve fits and updated orbital and planetary parameters. We discuss the implications of our phase curve fits for the planet's atmospheric dynamics in Section~\ref{sec:disc}.

\begin{table}[t!]
\small
	\centering
	\begin{threeparttable}
		\caption{Target System Properties} \label{tab:properties}	
		\renewcommand{\arraystretch}{1.2}
		\begin{center}
			\begin{tabular}{ l c c c c } 
				\hline\hline
				 &  WASP-19b & Ref. & HAT-P-7b  & Ref.\\
				 \hline
				 $T_{*}$~(K)  & $5568\pm71$& 3 & $6441\pm69$ & 3\\
				 $M_{*}$~($M_{\astrosun}$)  & $0.904\pm0.040$ & 2 & $1.47\pm0.07$ & 5\\
				 $R_{*}$~($R_{\astrosun}$)  & $1.004\pm 0.016$ & 2 & $1.84\pm0.17$ & 5\\
				 Stellar $\log g$ &  $4.45\pm0.05$ & 3 & $4.02\pm0.01$ & 3 \\
				 Stellar [Fe/H] &  $+0.15\pm0.07$ & 3 & $+0.15\pm0.08$ & 3\\
				 $M_{p}$~($M_{\mathrm{Jup}}$)  & $1.114\pm0.036$ & 2 & $1.776^{+0.077}_{-0.049}$ & 5\\
				 $R_{p}$~($R_{\mathrm{Jup}}$)  & $1.395 \pm 0.023$ & 2 & $1.36\pm0.15$ & 6\\
				 $P$~(d) & $0.78883942$ & 2 & $2.204737$ & 6\\
				  & $\qquad\quad\pm3.3$e-07  & & $\qquad\quad\pm1.7$e-05 &\\
				 $a$~(au)  & $0.0165^{+0.0005}_{-0.0006}$ & 1 & $0.0377\pm 0.0005$ & 5\\
				 $e$  & $0.0024^{+0.0094}_{-0.0019}$ & 4 & $0.0055^{+0.0070}_{-0.0033}$  & 4\\
				 $i$ &  $78.94\pm0.23$ & 2 & $83.111 \pm 0.030$ & 6\\
				 $\omega$~($^{\circ}$)  & $260^{+15}_{-170}$ & 4 & $204^{+53}_{-89}$ & 4 \\
				 $T_{0}$~(K)\textsuperscript{a}  & 3000 & ... & 3210 & ... \\
				 $T_{\mathrm{eq}}$~(K)\textsuperscript{b}  & 2520 &  ... & 2700  & ... \\
				\hline
			\end{tabular}
			
			\begin{tablenotes}
				\small
				\item {\bf Notes.} 
				\item \textsuperscript{a}Irradiation temperature at the substellar point assuming zero albedo: $T_{0}\equiv T_{*}\sqrt{R_{*}/a}$. 
				\item \textsuperscript{b}Planet dayside equilibrium temperature assuming zero albedo and reradiation from dayside only.
				\item {\bf References.} (1) \citet{hebb}, (2) \citet{tregloanreed}, (3) \citet{torres}, (4) \citet{knutson2014}, (5) \citet{pal}, (6) \citet{morris}.
				
			\end{tablenotes}
		\end{center}
	\end{threeparttable}
\end{table}

\section{\textit{Spitzer} observations and methods}\label{sec:obs}
For each of the two planets, we observed two full orbits: one orbit each in the 3.6 and 4.5~$\mu$m channels of the Infrared Array Camera \citep[IRAC;][]{fazio} on the Spitzer Space Telescope. All observations were carried out in subarray mode  with 2.0~s integration times and no peak-up pointing. Additional observation details can be found in Table~\ref{tab:obs}.

\begin{table*}[t!]
	\centering
	\begin{threeparttable}
		\caption{\textit{Spitzer} Observation Details} \label{tab:obs}	
		\renewcommand{\arraystretch}{1.2}
		\begin{center}
			\begin{tabular}{ l  c  c  c  c  c  c  c  c c  c } 
				\hline\hline
				Target &  $\lambda$~($\mu$m) &  UT Start Date &  Length~(hr) &  $n_{\mathrm{img}}$\textsuperscript{a} &  $t_{\mathrm{int}}$~(s)\textsuperscript{b}  & $t_{\mathrm{trim}}$~(hr)\textsuperscript{c}  & $r_{0}$\textsuperscript{c}  & $r_{1}$\textsuperscript{c}  & Fixed\textsuperscript{d} & $r_{\mathrm{phot}}$\textsuperscript{c} \\
				\hline
				WASP-19b &  3.6 &  2011 Aug 3 &  24.5 &  44352  & 2.0  & 0.5  & 3.0 & 1.0 & no & 1.55 \\
				  & 4.5  & 2011 Aug 13  & 24.5  & 44352   & 2.0  & 0.0  & 3.5 & 1.0 & no & 1.66 \\
				HAT-P-7b  & 3.6  & 2010 Aug 9  & 62.5  & 112000  & 2.0  & 0.5  & 4.0 & none & yes & 2.00\\
				  & 4.5  & 2010 Aug 20 & 62.5  & 112000   & 2.0  & 0.0  & 4.5 & 2.0 & no & 1.97 \\
				\hline

			\end{tabular}
			
			\begin{tablenotes}
				\small
				\item {\bf Notes.}
				\item \textsuperscript{a}Total number of images.
				\item \textsuperscript{b}Image integration time.
				\item \textsuperscript{c}$t_{\mathrm{trim}}$ is the amount of time trimmed from the start of each time series, $r_{0}$ is the radius of the aperture used to determine the star centroid position on the array, $r_{1}$ is the radius of the aperture used to compute the noise pixel parameter, and $r_{\mathrm{phot}}$ is the radius of the photometric aperture. We provide the median aperture radius over the observation in the case of a time-varying aperture. All radii are given in units of pixels. When using a fixed aperture, the noise pixel parameter is not needed, and so $r_{1}$ is undefined.
				\item \textsuperscript{d}Denotes whether the photometry was obtained using a fixed or time-varying aperture. 
				
			\end{tablenotes}
		\end{center}
	\end{threeparttable}
\end{table*}

\subsection{Photometry extraction}\label{subsec:photometry}

The techniques we use in extracting photometry are identical to those described in detail in several previous analyses of post-cryogenic \textit{Spitzer} data \citep[e.g.,][]{lewis,todorov,wong,wong2}. After the raw data are dark-subtracted, flat-fielded, linearized, and flux-calibrated using version S19.1.0 of the IRAC pipeline, we generate the BJD$_{\mathrm {UTC}}$ mid-exposure time series, subtract the sky background, and correct for hot pixels using the methods described in \citet{knutson2012} and \citet{lewis}.

\begin{figure}[b]
\begin{center}
\includegraphics[width=9cm]{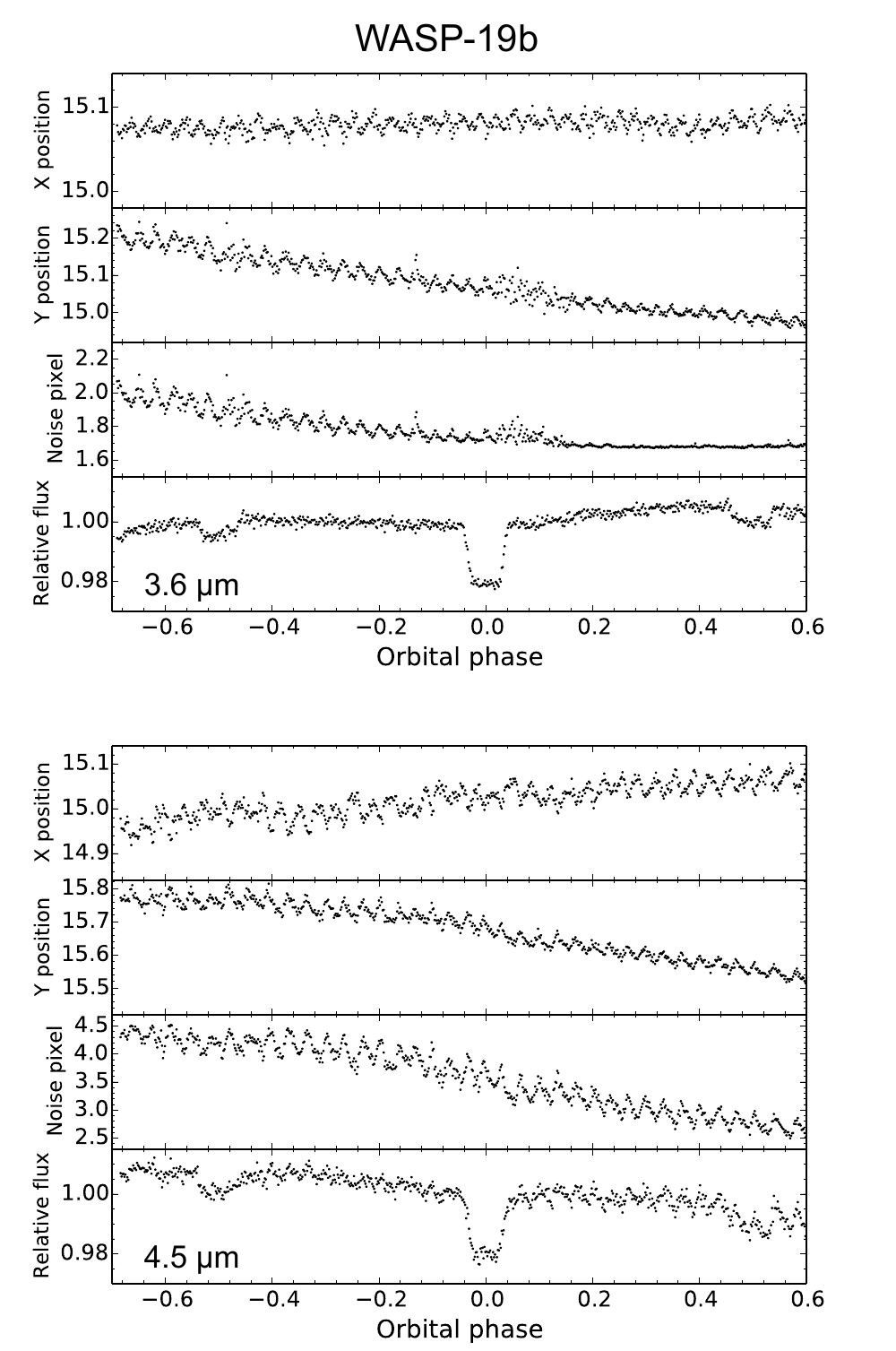}
\end{center}
\caption{Measured stellar $x$ centroids, $y$ centroids, noise pixel values, and raw photometric series with hot pixels excised as a function of orbital phase relative to transit for the 3.6~$\mu$m (top) and 4.5~$\mu$m (bottom) phase curve observations of WASP-19b. The data are binned in two-minute intervals.} \label{data1}
\end{figure}

\begin{figure}[t]
\begin{center}
\includegraphics[width=9cm]{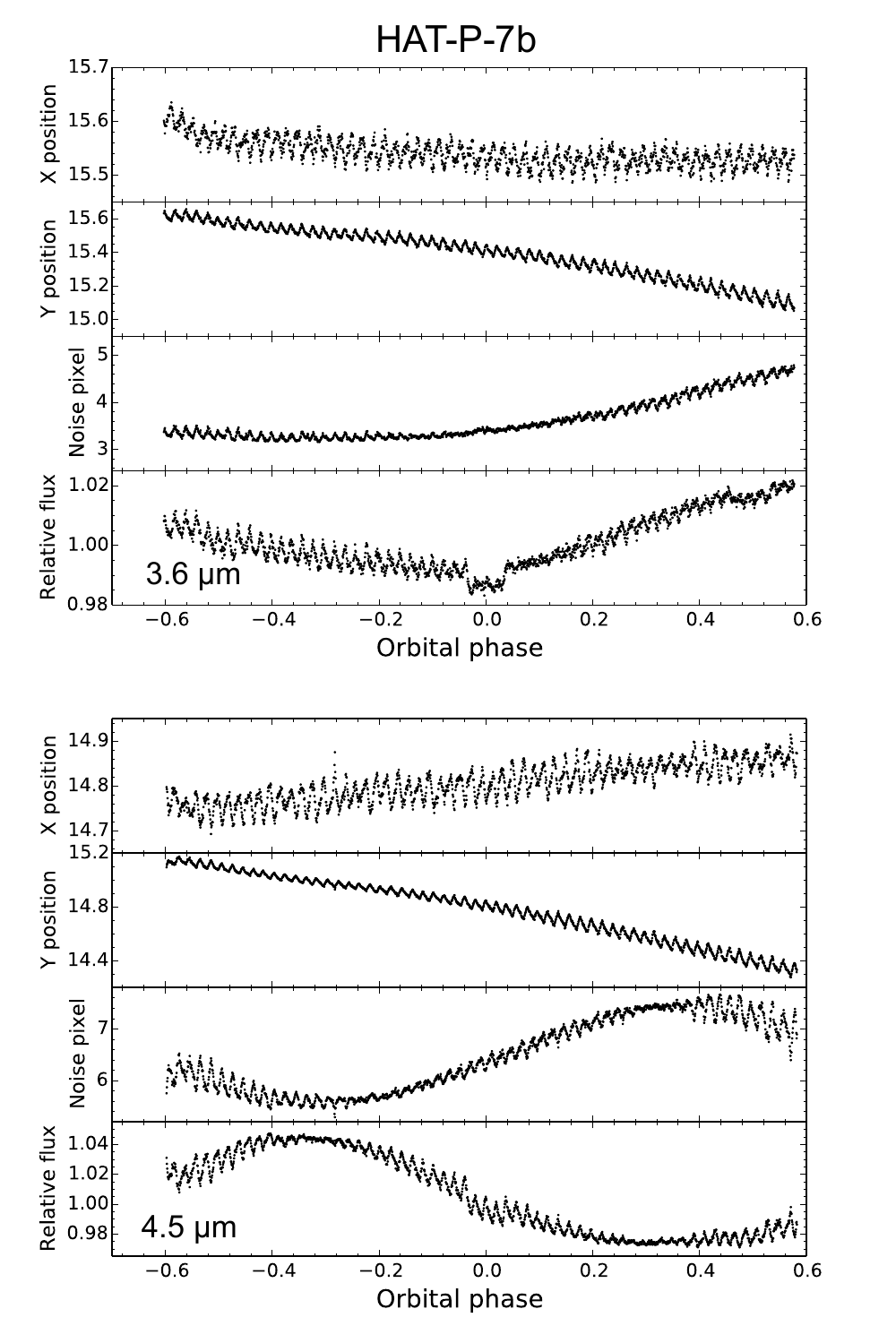}
\end{center}
\caption{Measured star centroids, noise pixel values, and raw photometric series for the HAT-P-7b phase curve observations; see Figure~\ref{data1} for a complete description.} \label{data2}
\end{figure}

We calculate the position of the star in each image using flux-weighted centroiding over a circular region of radius $r_{0}$ pixels centered on the approximate position of the star on the array \citep[see, for example,][]{knutson2008}. The width of the star's point response function, which is the convolution of the star's point-spread function (PSF) and the detector response function, is estimated by computing the noise pixel parameter $\widetilde{\beta}$ \citep{mighell} across a region of radius $r_{1}$  centered on the star centroid.

We obtain the flux of the stellar target using both fixed and time-varying circular apertures where, in the case of time-varying apertures, the radius is related to the square-root of the noise pixel parameter $\widetilde{\beta}$ with either a constant scaling factor or a constant shift \citep{lewis}.  Each set of fixed and time-varying apertures is generated for several choices of $r_{0}$ and $r_{1}$ (for time-varying apertures only). For every phase curve observation, we determine the optimal aperture by selecting the aperture that produces the least scatter in the residual series from the phase curve model fit, binned in five-minute intervals. When selecting the optimal aperture, we fix the planets' period, orbital eccentricity, and argument of pericenter to the most recent values in the literature (see Table~\ref{tab:properties}). 

For WASP-19b, the optimal aperture for the 3.6~$\mu$m data set has a time-varying radius of $1.35\sqrt{\widetilde{\beta}}$, $r_{0}=3.0$, and $r_{1}=1.0$, while for the 4.5~$\mu$m data set, we prefer a time-vaying aperture with a radius of $\sqrt{\widetilde{\beta}}+1.2$, $r_{0}=3.5$, and $r_{1}=1.0$. For the HAT-P-7b phase curves, we find that a fixed aperture of radius 2.0~pixels with $r_{0}=4.0$ results in the lowest scatter in the 3.6~$\mu$m bandpass, while in the 4.5~$\mu$m bandpass, we prefer a time-varying aperture with a radius of $1.45\sqrt{\widetilde{\beta}}$, $r_{0}=4.5$, and $r_{1}=2.0$. The relevant information describing our optimal apertures is summarized in Table~\ref{tab:obs}. Figures~\ref{data1} and \ref{data2} show the measured star positions, noise pixel values, and extracted hot pixel-corrected photometric series for the 3.6 and 4.5~$\mu$m phase curve observations of WASP-19b and HAT-P-7b, respectively.

For each photometric series, we apply a moving median filter to the flux, $x$ and $y$ star centroid position, and $\sqrt{\widetilde{\beta}}$ series and iteratively remove points that vary by more than 3$\sigma$ from the corresponding median values in the adjacent 64 points in the time series. For the WASP-19b observations, the percentages of excised points are 3.3\% and 2.6\% in the 3.6 and 4.5~$\mu$m bandpasses, respectively, while for the HAT-P-7b observations, the corresponding percentages are 1.9\% and 1.4\%.

\subsection{Full phase curve model}\label{subsec:phasemodel}
In this work, we use the same phase curve model as in \citet{wong2}. Both WASP-19b and HAT-P-7b have nearly circular orbits, and so we model the planet's apparent brightness modulation throughout an orbit as a sinusoidal function of the true anomaly $f$ \citep{lewis},
 \begin{equation}\label{phase}F(t)=F_{0}(t)+c_{1}\cos(f(t)-c_{2}), \end{equation}
where $F_{0}$ is the star's flux, $c_{1}$ is the amplitude of the phase variation, and $c_{2}$ is the phase shift. We set $c_{1}$ and $c_{2}$ as free parameters in our fits. We considered including higher harmonics in our phase curve model but found that they resulted in higher values of the Bayesian Information Criterion (BIC; see Section~\ref{subsec:stellar}). In light of the low orbital eccentricities of both planets, we also experimented with setting the eccentricity to zero in our global phase curve fits. The best-fit phase curve parameters in the zero-eccentricity fits are consistent  at better than the $0.2\sigma$ level with the corresponding values computed in the fits with eccentricities derived from our updated RV analysis (see Section~\ref{sec:analysis}). In this paper, we report the best-fit parameter values derived from our fits with non-zero eccentricity.

Each phase curve observation contains one transit and two secondary eclipses, which we model using the formalism of \citet{mandelagol}. In our full phase curve analysis, we fit for the scaled orbital semi-major axis $a/R_{*}$, the inclination $i$, the center of transit time $t_{T}$, and the planet--star radius ratio $R_{p}/R_{*}$, as well as the center of eclipse time $t_{E}$ and the relative eclipse depth $d$ of each secondary eclipse. In the final version of our phase curve fits, we fix the orbital eccentricity $e$, argument of perihelion $\omega$, and orbital period $P$ to the updated values obtained from our RV and ephemeris analysis (see Section~\ref{sec:analysis}). We model the effect of limb-darkening on the transit light curve in each bandpass using the four-parameter nonlinear limb-darkening law described in \citet{sing}. The relevant host star properties are listed in Table~\ref{tab:properties}. For WASP-19, we use the limb-darkening parameter values calculated for a 5500~K star with  $\log{g}=4.50$ and [Fe/H] $= +0.10$: $c_{1}-c_{4}=[0.5437,-0.4959,0.4787,-0.1852]$ at 3.6~$\mu$m and $c_{1}-c_{4}=[0.6287,-0.8565,0.8561,-0.3164]$ at 4.5~$\mu$m.\footnote[1]{Tables of limb-darkening parameter values calculated in the \textit{Spitzer} bandpasses are provided on David Sing's website: www.astro.ex.ac.uk/people/sing} For HAT-P-7b, we take the limb-darkening parameters calculated for a 6500~K star with  $\log{g}=4.00$ and [Fe/H] $= +0.10$: $c_{1}-c_{4}=[0.4972,-0.5359,0.5630,-0.2367]$ at 3.6~$\mu$m and $c_{1}-c_{4}=[0.5601,-0.8206,0.8745,-0.3416]$ at 4.5~$\mu$m.

\subsection{Stellar variability}\label{subsec:stellar}

WASP-19 is an active G-type star with a measured rotation period of $10.7\pm 0.5$~days and photometric variability due to starspots at the $1.8\%$ (peak-to-peak) level in the visible \citep{abe}. Although the star's rotation period is much longer than the duration of our phase curve measurements, and the level of stellar variability in the infrared is reduced relative to its amplitude in the visible, the stellar contribution to light curve variations is still comparable to the amplitude of the planet's phase curve and must be accounted for. There was no simultaneous monitoring of the host star during our phase curve observations. We therefore model the effects of stellar variability in our data as a quadratic function of time \citep{knutson2012}:
\begin{equation}\label{star}F_{0}(t) = \epsilon_{1}(t-t_{T})+\epsilon_{2}(t-t_{T})^2,\end{equation}
where $\epsilon_{1}$ and $\epsilon_{2}$ are free parameters. To determine how many terms are necessary for modeling the stellar variability in each phase curve data set, we use the BIC, which is defined as
\begin{equation}\label{bic}\mathrm{BIC} = \chi^2 + k\ln{N},\end{equation}
where $k$ is the total number of free parameters in the fit, and $N$ is the number of data points. We select the stellar variability model that minimizes the BIC. For the 3.6~$\mu$m WASP-19b data, we find that including both terms yields a significantly lower BIC compared to the case with just a linear term, while for the 4.5~$\mu$m data, a linear stellar variability function results in the lowest BIC. The addition of higher-order terms does not improve the BIC in either bandpass. For completeness, we also experimented with including stellar variability in our HAT-P-7b phase curve fits, but did not find a reduction in BIC.

\subsection{Correction for instrumental effects}\label{subsec:correction}
Small oscillations in the telescope pointing during \textit{Spitzer}/IRAC observations in the 3.6 and 4.5~$\mu$m bandpasses lead to variations in the measured flux from the target due to intrapixel sensitivity variations \citep{charbonneau2005}. In our analysis, we utilize two methods for decorrelating this instrumental systematic: pixel mapping \citep[][]{ballard,lewis} and pixel-level decorrelation \citep[PLD; see][for a complete description]{pld}.

\citet{pld} found that results from PLD were comparable to or better than those from other methods as long as the range of star positions across the data set remains below $\sim$0.2~pixels. In our full phase curve observations, the ranges of star centroids exceed the 0.2~pixel threshold (see Figures~\ref{data1} and \ref{data2}). In \citet{wong2}, we compared the performance of pixel mapping and PLD in full phase curve fits and determined that pixel mapping consistently produced residuals with lower variance. Therefore, we use pixel mapping when fitting the full phase curve data sets in this work. On the other hand, when carrying out individual fits of HAT-P-7b secondary eclipses (see Section~\ref{sec:analysis}), where there is significant residual red noise in the light curve after decorrelating the intrapixel sensitivity variations with the pixel mapping technique and the range of star positions is under 0.2~pixels, we find that PLD is more effective than pixel mapping at reducing the level of residual noise.

In addition to the intrapixel sensitivity effect, data obtained using \textit{Spitzer}/IRAC often exhibit a short-duration ramp at the beginning of the observation \citep[e.g.][]{knutson2012,lewis}. We first experimented with removing the first 30, 60, or 90 minutes of data from each phase curve observation and found that a removal interval of 30 minutes yields the minimum binned residual RMS  for the 3.6~$\mu$m observations; meanwhile, for the 4.5~$\mu$m observations, the binned residual RMS is minimized without removing any data. 

Next, we considered adding a ramp function to our phase curve model to further improve our fits. We model the ramp as an exponential function of time \citep{agol}:
\begin{equation}\label{ramp}F=1\pm a_{1}\mathrm{e}^{-(t-t_{0})/a_{2}}\pm a_{3}\mathrm{e}^{-(t-t_{0})/a_{4}},\end{equation}
where $t_{0}$ is the observation start time, and $a_{1}$---$a_{4}$ are free parameters. We determine the number of exponential terms to include in the ramp model by means of the BIC. For the WASP-19b phase curve observations, we find that the BIC is minimized when using a single exponential ramp in the 3.6~$\mu$m bandpass; a double ramp is needed for the 4.5~$\mu$m data. For the HAT-P-7b data, we prefer a single exponential ramp at 3.6~$\mu$m and no ramp at 4.5~$\mu$m. In addition to exponential ramps, we experimented with ramps of different forms, such as logarithmic and quadratic; we find that these other forms result in fits with residual scatter and BIC values comparable to or larger than those obtained by fitting exponential ramps.

\section{Results}\label{sec:analysis}

\begin{figure*}[b]
\begin{center}
\includegraphics[width=17cm]{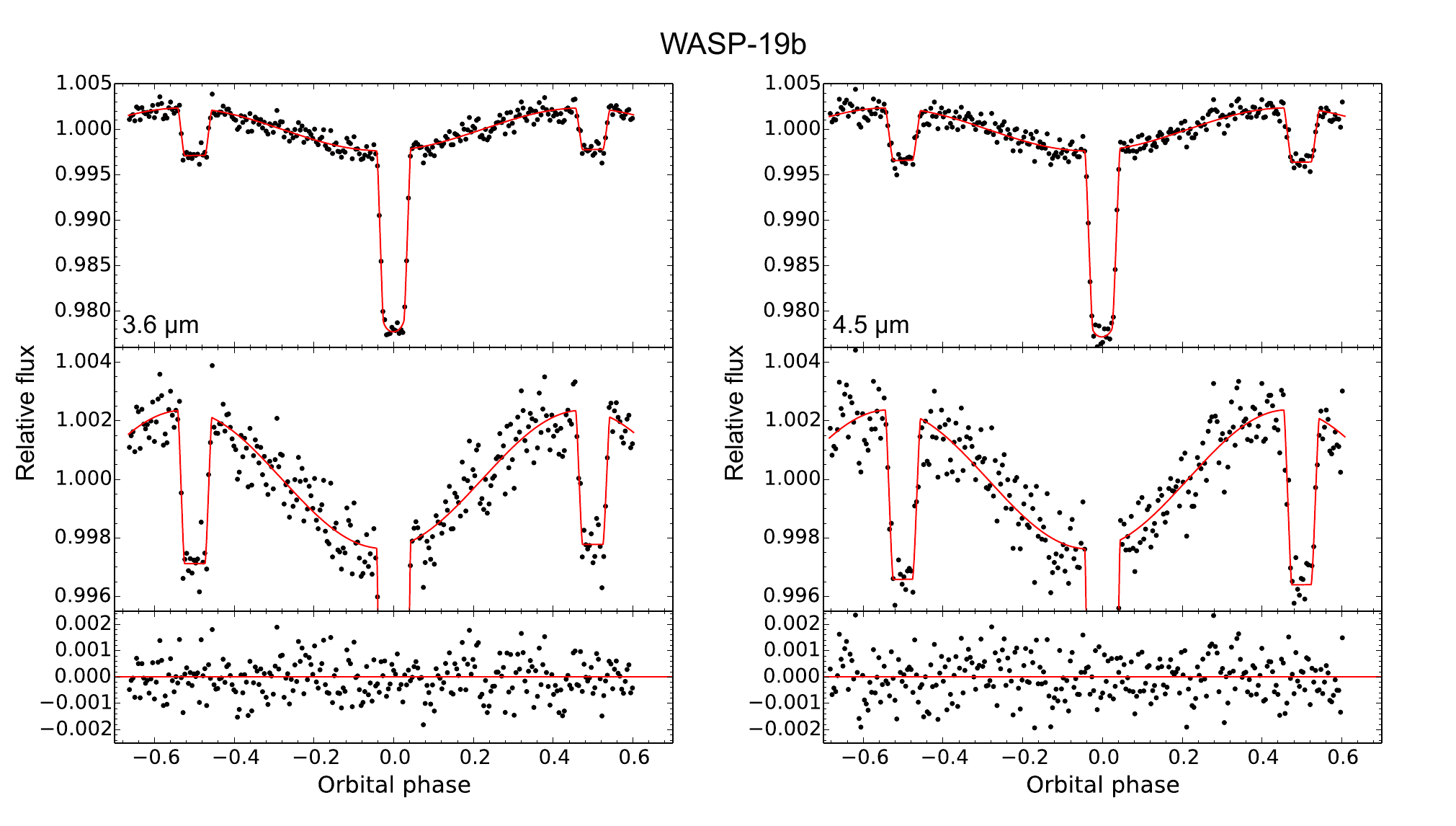}
\end{center}
\caption{Top panels: final photometric series for the 3.6 and 4.5~$\mu$m WASP-19b phase curve observations with instrumental systematics removed, binned in five-minute intervals (black dots). The best-fit single and double exponential ramp models at 3.6 and 4.5~$\mu$m, respectively, are also removed. The best-fit model light curve is overplotted in red. Middle panels: the same data as the upper panel, but with y axis expanded for clarity. Bottom panels: the residuals from the best-fit solution.}\label{phase1}
\end{figure*}

\begin{figure*}[t]
\begin{center}
\includegraphics[width=17cm]{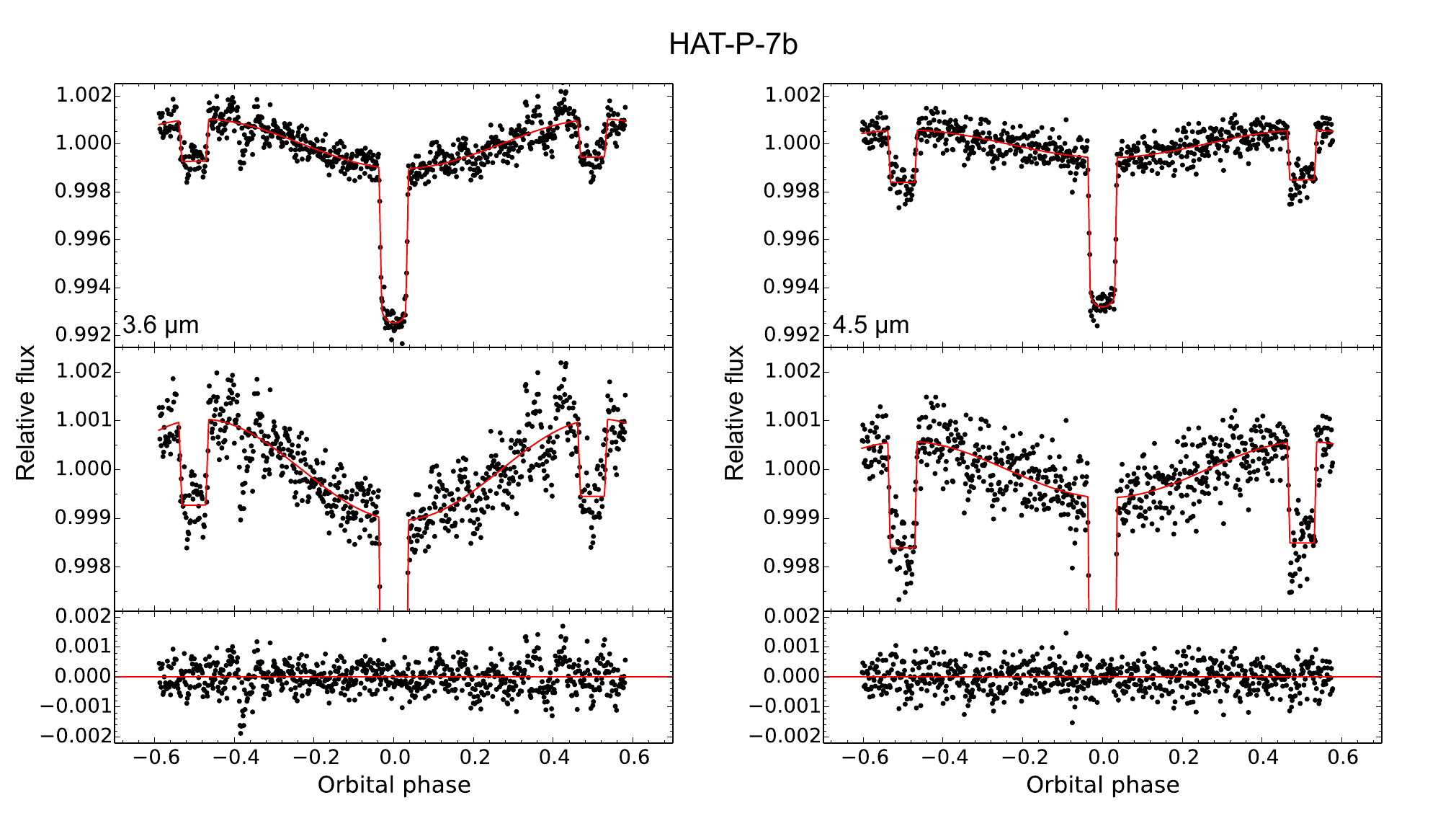}
\end{center}
\caption{Final photometric series for the HAT-P-7b phase curve observations with instrumental systematics removed. The best-fit single exponential ramp model is removed from the 3.6~$\mu$m light curve data; no ramp is used in the 4.5~$\mu$m phase curve fit. See Figure~\ref{phase1} for a complete description.} \label{phase2}
\end{figure*}

\begin{figure}[t]
\begin{center}
\includegraphics[width=9cm]{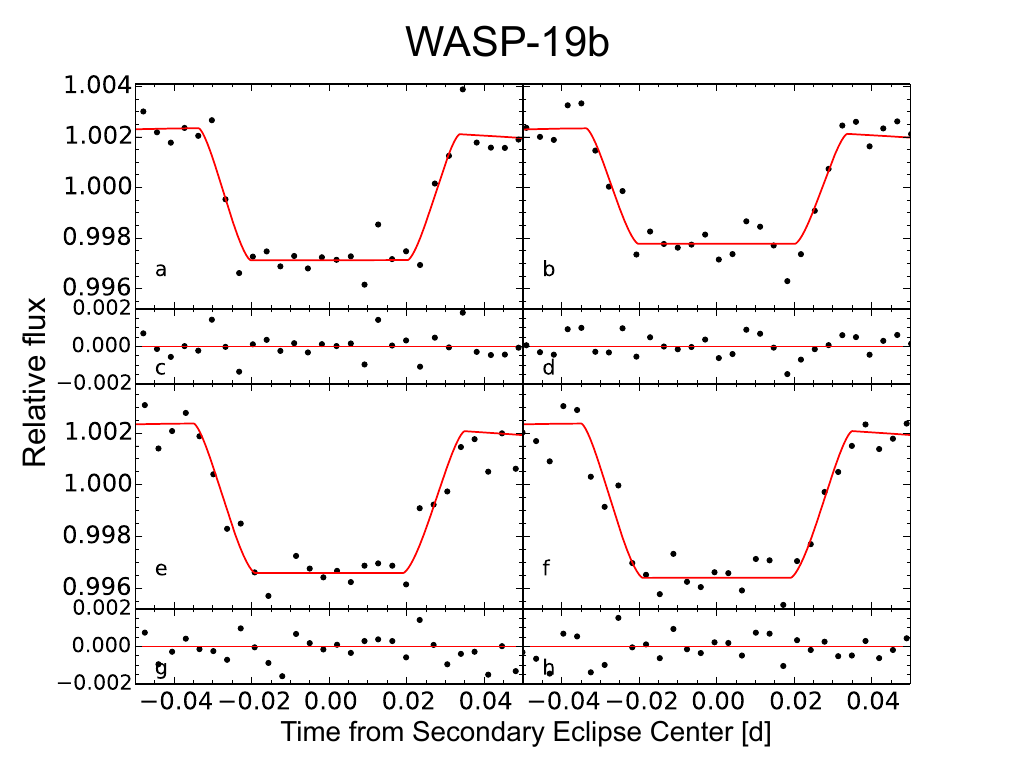}
\end{center}
\caption{Best-fit WASP-19b eclipse light curve data in the 3.6~$\mu$m (a$-$d) and 4.5~$\mu$m (e$-$h) bands with intrapixel sensitivity variations removed, binned in five-minute intervals (black dots). The best-fit model light curves are overplotted in red. The residuals from the best-fit solution (c$-$d, g$-$h) are shown directly below the corresponding light curve data (a$-$b; e$-$f).} \label{eclipses1}
\end{figure}

\begin{figure}[t]
\begin{center}
\includegraphics[width=9cm]{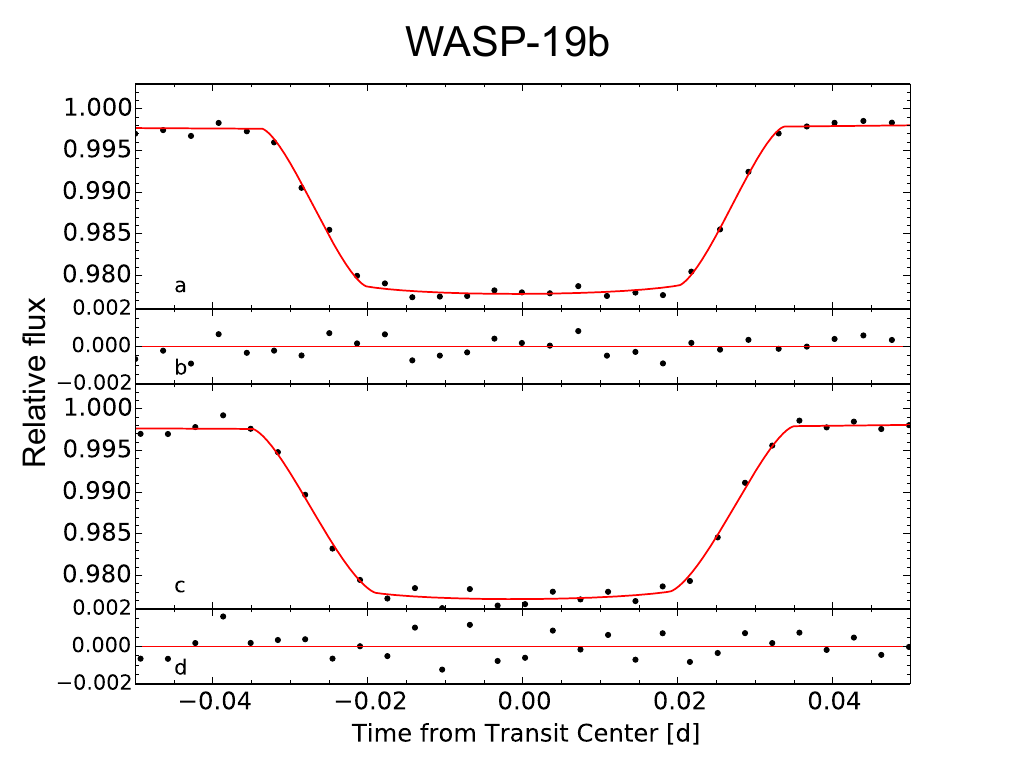}
\end{center}
\caption{Best-fit WASP-19b transit light curve data in the 3.6~$\mu$m and 4.5~$\mu$m bands with intrapixel sensitivity variations removed, binned in five-minute intervals (black dots).  The best-fit model light curves are overplotted in red. The corresponding residuals from the best-fit solution are shown directly below each transit light curve.} \label{transits1}
\end{figure}

\begin{figure}[t]
\begin{center}
\includegraphics[width=9cm]{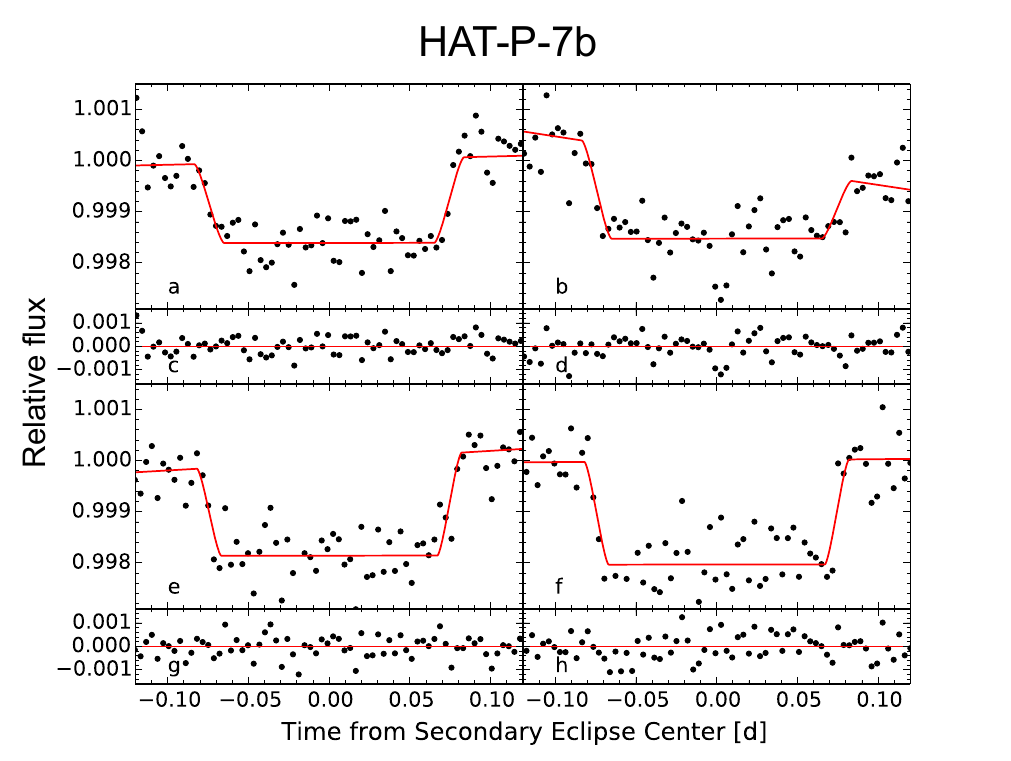}
\end{center}
\caption{Best-fit HAT-P-7b eclipse light curves with intrapixel sensitivity variations removed, binned in five-minute intervals (black dots), and the best-fit model light curves (red lines).  These light curves are derived from fitting each eclipse event separately. See Figure~\ref{eclipses1} for a complete description of the panels.} \label{eclipses2}
\end{figure}

\begin{figure}[t]
\begin{center}
\includegraphics[width=9cm]{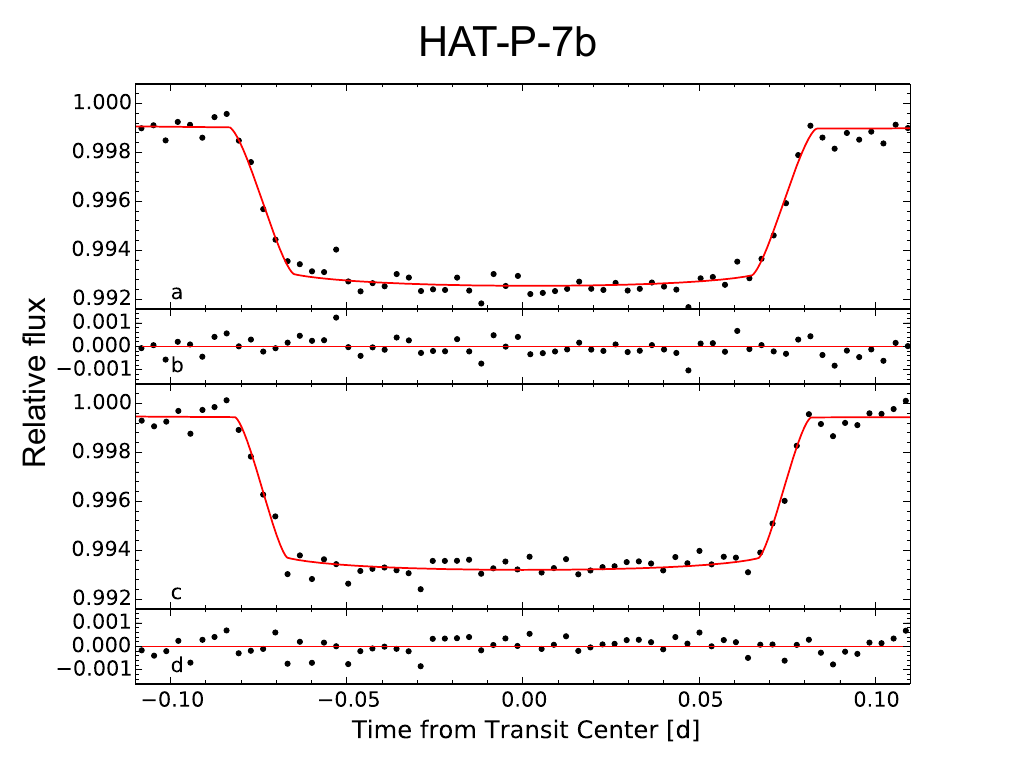}
\end{center}
\caption{Best-fit HAT-P-7b transit light curves with intrapixel sensitivity variations removed, binned in five-minute intervals (black dots), and the best-fit model light curves (red lines). See Figure~\ref{transits1} for a complete description.} \label{transits2}
\end{figure}

The best-fit parameter values and uncertainties derived from our  WASP-19b and HAT-P-7b phase curve fits are listed in Tables~\ref{tab:values1} and \ref{tab:values2}, respectively. As in \citet{wong2}, we calculate the uncertainties  using both the ``prayer-bead" (PB) method of residual permutation \citep{gillon} and a Markov chain Monte Carlo (MCMC) routine with $10^5$ steps, reporting the larger of the two errors for each parameter. The PB uncertainties are usually larger, ranging between 0.9 and 3.6 times that of the corresponding MCMC uncertainties for the phase curve parameters. The binned full-orbit photometric series with instrumental variations and ramps removed are shown in Figures~\ref{phase1} and \ref{phase2}, while the individual eclipse and transit light curves are shown in Figures~\ref{eclipses1}$-$\ref{transits2}.

\begin{figure}[t]
\begin{center}
\includegraphics[width=8cm]{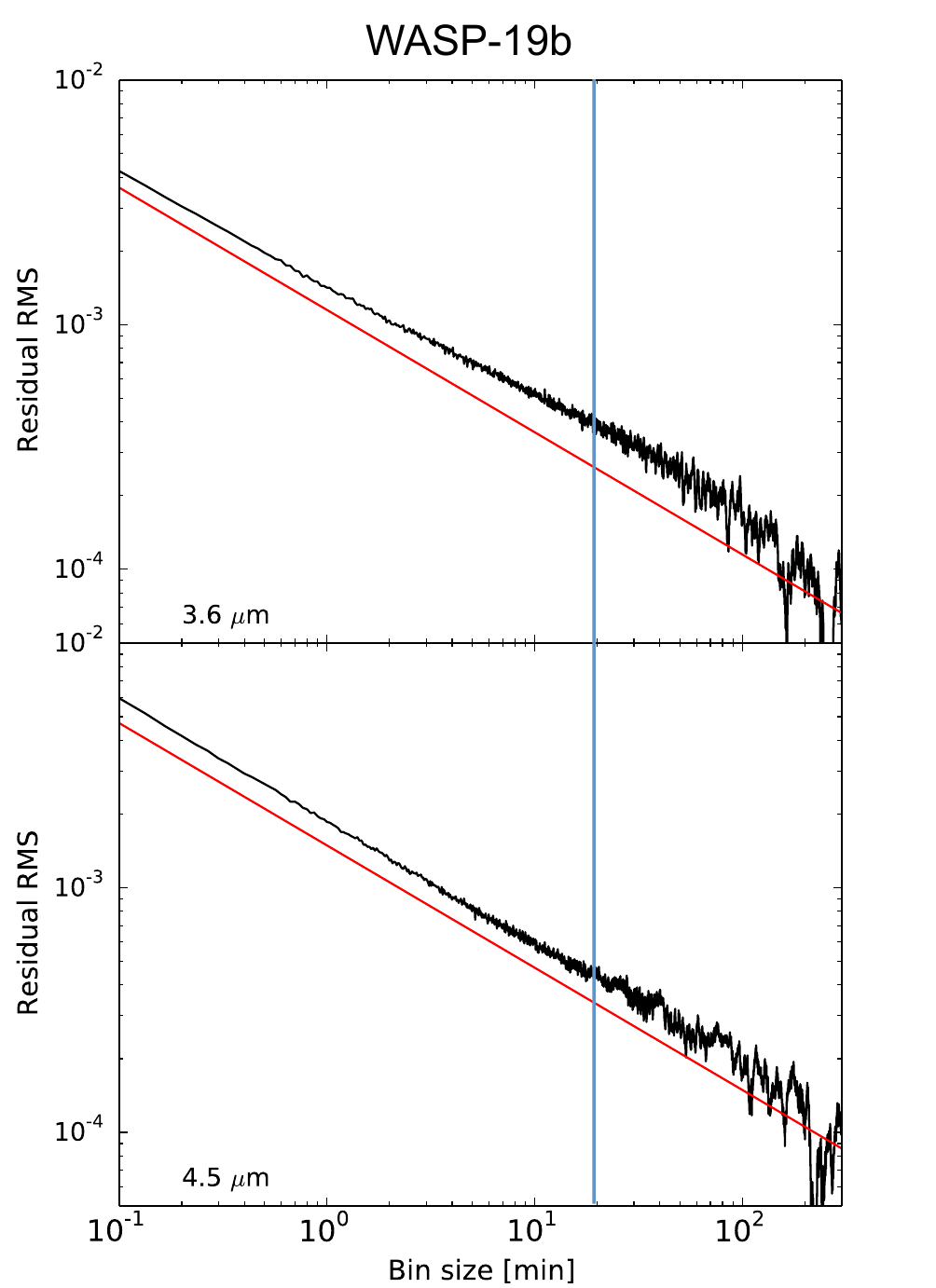}
\end{center}
\caption{Plot showing the dependence of the standard deviation of the residuals on bin size for the 3.6~$\mu$m (top panel) and 4.5~$\mu$m (bottom panel) WASP-19b data sets (black lines). The $1/\sqrt{n}$ dependence of white noise on bin size is shown by the red lines for comparison. The normalization of the white noise trends are set to match the expected (unbinned) photon noise limit. The blue vertical line denotes the ingress/egress timescale of transit and secondary eclipse.}\label{rednoise1}
\end{figure}

\begin{figure}[t]
\begin{center}
\includegraphics[width=8cm]{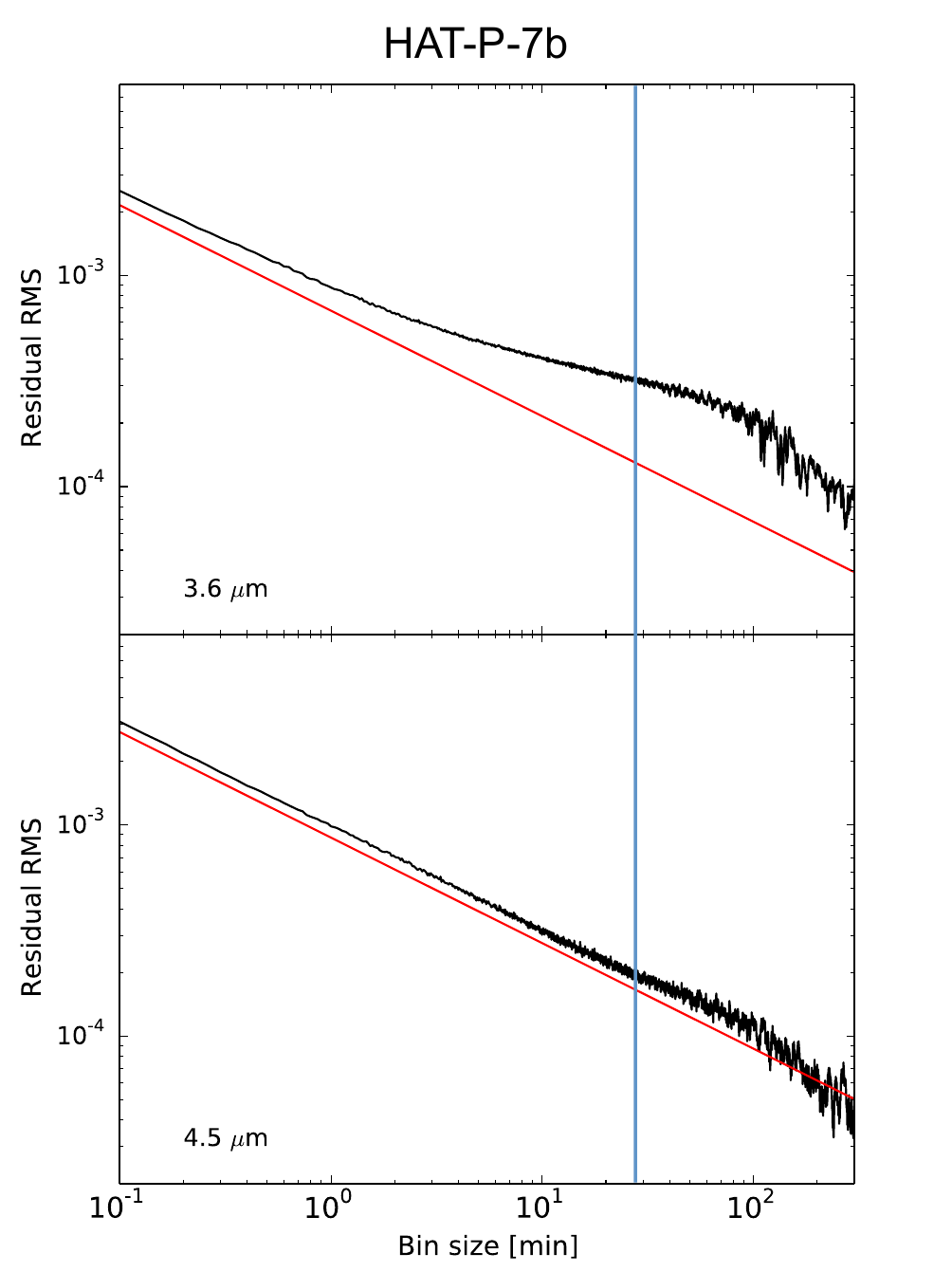}
\end{center}
\caption{Plot of the standard deviation of the residuals versus bin size for the 3.6~$\mu$m (top panel) and 4.5~$\mu$m (bottom panel) HAT-P-7b data sets. See Figure~\ref{rednoise1} for a complete description.}\label{rednoise2}
\end{figure}

The residual RMS scatter is 15\% and 28\% (WASP-19b) and 14\% and 12\% (HAT-P-7b) higher than the photon noise limit in the 3.6 and 4.5~$\mu$m bands, respectively. Figures~\ref{rednoise1} and \ref{rednoise2} show the residual RMS from the best-fit solution for various bin sizes. Comparing these values with the expected level of white noise, we find that on timescales comparable to the transit ingress/egress duration, the red noise increases the RMS by factors of roughly 1.6 and 1.3 (WASP-19b) and 2.5 and 1.2 (HAT-P-7b) at 3.6 and 4.5~$\mu$m, respectively.

\begin{table}[t!]
\scriptsize
\centering
\begin{threeparttable}
\caption{WASP-19\MakeLowercase{b} Best-fit Parameters} \label{tab:values1}

\renewcommand{\arraystretch}{1.2}
\begin{center}
\begin{tabular}{ l  r r }
\hline\hline
 Parameter & 3.6~$\mu$m & 4.5~$\mu$m \\
\hline
\textit{Transit Parameters} & &\\
$R_{p}/R_{*}$ & $0.1399^{+0.0014}_{-0.0018}$ &  $0.1427^{+0.0017}_{-0.0025}$ \\
$t_{T}$~(BJD$-2455770$)\textsuperscript{a} & $7.16289^{+0.00022}_{-0.00021}$ &  $17.41786^{+0.00023}_{-0.00022}$ \\

\\
\textit{Eclipse Parameters} & &\\
1\ts{st} eclipse depth, $d_{1}$~($\%$) & $0.521^{+0.038}_{-0.035}$ & $0.575^{+0.044}_{-0.039}$ \\
$t_{E1}$~(BJD$-2455770$)\textsuperscript{a}\textsuperscript{,}\textsuperscript{b} & $6.76944^{+0.00082}_{-0.00083}$ &  $17.02306^{+0.00078}_{-0.00076}$ \\
2\ts{nd} eclipse depth, $d_{2}$~($\%$) & $0.455^{+0.035}_{-0.031}$ & $0.593^{+0.040}_{-0.039}$ \\
$t_{E2}$~(BJD$-2455770$)\textsuperscript{a}\textsuperscript{,}\textsuperscript{b} & $7.55656^{+0.00095}_{-0.00088}$ & $17.81151^{+0.00077}_{-0.00074}$ \\

\\
\textit{Orbital Parameters} & &\\
Inclination, $i$~($^{\circ}$) & $79.63^{+0.84}_{-0.82}$ & $77.98^{+0.78}_{-0.84}$\\
Scaled semi-major axis,  & $3.59^{+0.12}_{-0.11}$ & $3.36\pm 0.10$ \\
\qquad $a/R_{*}$ & & \\
\\
\textit{Phase Curve Parameters} & &\\
Amplitude, $c_{1}$~($\times 10^{-3}$) &  $2.36^{+0.17}_{-0.18}$ &  $2.37^{+0.23}_{-0.20}$ \\
Phase shift, $c_{2}$~($^{\circ}$) &  $0.6^{+4.6}_{-6.6}$ &   $-2.2^{+4.5}_{-5.6}$ \\
Maximum flux offset~(h)\textsuperscript{b} &  $-0.55\pm0.21$ &   $-0.68\pm0.19$ \\
Minimum flux offset~(h)\textsuperscript{b} & $-0.55^{+0.24}_{-0.35}$ &   $-0.70^{+0.24}_{-0.30}$ \\

\\
\textit{Ramp/Stellar Parameters}\textsuperscript{d}  & &\\
$a_{1}$ ($\times 10^{-2}$) & $1.92^{+4.73}_{-0.40}$  & $0.26^{+1.74}_{-0.15}$ \\
$a_{2}$ (d)&  $0.38^{+0.49}_{-0.18}$ & $0.11^{+0.55}_{-0.10}$ \\
$a_{3}$ ($\times 10^{-2}$) & 0 (fixed)  & $0.32^{+2.21}_{-1.00}$ \\
$a_{4}$ (d)& 0 (fixed) &  $0.95^{+2.12}_{-0.65}$ \\
$\epsilon_{1}$ ($\times 10^{-2}$ d$^{-1}$) & $-1.26^{+0.31}_{-3.53}$ & $-0.41^{+0.34}_{-1.07}$ \\
$\epsilon_{2}$ ($\times 10^{-2}$ d$^{-2}$) & $1.13^{+2.37}_{-0.77}$ & 0 (fixed) \\

\hline
\end{tabular}

\begin{tablenotes}
      \footnotesize
      \item {\bf Notes.}
      \item \textsuperscript{a}All times are listed in BJD$_{\mathrm {UTC}}$ for consistency with other studies. To convert to BJD$_{\mathrm {TDB}}$, add 65.184~s. The center of secondary eclipse times are not corrected for the light travel time across the system ($\Delta t = 16.3$~s).
      \item \textsuperscript{b}The maximum and minimum flux offsets are measured relative to the center of secondary eclipse time and center of transit time, respectively, and are derived from the phase curve fit parameters $c_{1}$ and $c_{2}$. The phase curve is modeled as a single sinusoid as a function of the true anomaly $f$: $F=1+c_{1}\cos(f-c_{2})$.
      \item \textsuperscript{c}The exponential ramp at the beginning of observation is parametrized as $F(t) = -a_{1}\exp{\left\lbrack(t-t_{0})/a_{2}\right\rbrack}-a_{3}\exp{\left\lbrack(t-t_{0})/a_{4}\right\rbrack}$, where $t_{0}$ is the observation start time. The variation in stellar brightness is modeled as a second-order function in time: $F(t)=1+\epsilon_{1}(t-t_{T})+\epsilon_{2}(t-t_{T})^{2}$.

    \end{tablenotes}
    \end{center}
    \end{threeparttable}
\end{table}

\begin{table}[t!]
\scriptsize
\centering
\begin{threeparttable}
\caption{HAT-P-7\MakeLowercase{b} Best-fit Parameters} \label{tab:values2}

\renewcommand{\arraystretch}{1.2}
\begin{center}
\begin{tabular}{l r r}
\hline\hline
Parameter& 3.6~$\mu$m & 4.5~$\mu$m \\
\hline
\textit{Transit Parameters}  & &\\
$R_{p}/R_{*}$ & $0.0793^{+0.0015}_{-0.0012}$ &  $0.07769^{+0.00076}_{-0.00080}$ \\
$t_{T}$~(BJD$-2455410$)\textsuperscript{a} & $9.55743^{+0.00054}_{-0.00070}$ & $20.58203^{+0.00040}_{-0.00047}$ \\

\\
\textit{Eclipse Parameters}\textsuperscript{b} & &\\
1\ts{st} eclipse depth, $d_{1}$~($\%$) & $0.161^{+0.014}_{-0.015}$ & $0.186^{+0.008}_{-0.006}$ \\
$t_{E1}$~(BJD$-2455410$)\textsuperscript{a} & $8.4554^{+0.0021}_{-0.0019}$  & $19.4772^{+0.0010}_{-0.0009}$ \\
2\ts{nd} eclipse depth, $d_{2}$~($\%$) & $0.153^{+0.011}_{-0.010}$ & $0.204^{+0.014}_{-0.013}$ \\
$t_{E2}$~(BJD$-2455410$)\textsuperscript{a} & $10.6591^{+0.0022}_{-0.0018}$ & $21.6849\pm0.0012$ \\

\\
\textit{Orbital Parameters} & &\\
Inclination, $i$~($^{\circ}$) & $81.3\pm1.5$ & $83.7^{+2.2}_{-1.6}$ \\
Scaled semi-major axis, $a/R_{*}$ & $3.88^{+0.22}_{-0.21}$ & $4.23^{+0.28}_{-0.23}$ \\

\\
\textit{Phase Curve Parameters} & &\\
Amplitude, $c_{1}$~($\times 10^{-4}$) & $10.4^{+1.7}_{-1.8}$ &  $5.8^{+0.9}_{-1.0}$ \\
Phase shift, $c_{2}$~($^{\circ}$) & $112\pm 11$ & $109^{+10}_{-11}$ \\
Maximum flux offset~(h)\textsuperscript{c} & $1.0\pm 1.1$ & $0.6\pm 1.1$ \\
Minimum flux offset~(h)\textsuperscript{c} & $1.1^{+1.5}_{-1.6}$ & $0.6^{+1.5}_{-1.7}$ \\

\\
\textit{Ramp Parameters}\textsuperscript{d} & &\\
$a_{1}$ ($\times 10^{-3}$) &  $2.38^{+0.54}_{-0.49}$ &   --- \\
$a_{2}$ (d) &  $0.069^{+0.039}_{-0.023}$ &   --- \\

\hline
\end{tabular}

\begin{tablenotes}
      \footnotesize
       \item {\bf Notes.}
      \item \textsuperscript{a}All times are listed in BJD$_{\mathrm{UTC}}$ for consistency with other studies. To convert to BJD$_{\mathrm{TDB}}$, add 65.184 seconds. The center of secondary eclipse times are not corrected for the light travel time across the system ($\Delta t = 36.7$~s).
      \item \textsuperscript{b}These values are computed from fitting each secondary eclipse separately.
      \item \textsuperscript{c}The maximum and minimum flux offsets are measured relative to the center of secondary eclipse time and center of transit time, respectively, and are derived from the phase curve fit parameters $c_{1}$ and $c_{2}$.
      \item \textsuperscript{d}The exponential ramp at the beginning of observation is parametrized as $F(t) = -a_{1}\exp{\left\lbrack(t-t_{0})/a_{2}\right\rbrack}$, where $t_{0}$ is the observation start time.

    \end{tablenotes}
    \end{center}
    \end{threeparttable}
\end{table}

In our global fits, the full light curve model includes models for the exponential ramp, stellar variability, and the intrapixel sensitivity effect, in addition to the phase curve model. In order to check that the best-fit parameters of the phase curve model are unique for each measured phase curve and not biased by correlations with instrumental systematics and/or stellar variability, we carried out a global search over a two-dimensional grid of values for the phase curve amplitude $c_{1}$ and phase offset $c_{2}$. At each point within this $(c_{1},c_{2})$ grid, we computed the best-fit light curve model by minimizing $\chi^{2}$ over all other parameters, while keeping $(c_{1},c_{2})$ fixed at the value for that grid point. For each phase curve, we found that a unique minimum $\chi^{2}$ exists in this parameter space and lies at the grid point that is most consistent with the corresponding best-fit $(c_{1},c_{2})$ values computed from the full global fit (with $c_{1}$ and $c_{2}$ as free parameters; Tables~\ref{tab:values1} and \ref{tab:values2}). This shows that there are no significant degeneracies between the phase curve model and our models for instrumental systematics and stellar variability.

The intrapixel sensitivity-corrected phase curves of HAT-P-7b shown in Figure~\ref{phase2} indicate a significant level of uncorrected instrumental noise, particularly at 3.6~$\mu$m. This may be attributable to the significant long-term drift of the stellar target across the array (Figure~\ref{data2}). A wider range of star positions means that the density of sampling in any given region of the pixel is lower, resulting in larger uncertainties when calculating the pixel response at a given position. Furthermore, in the 3.6~$\mu$m data set, the $y$ position of the stellar target crosses over a pixel boundary, which corresponds to the notable dip in the residuals at roughly $-0.4$ in orbital phase. We experimented with trimming points with $y$ positions within 0.05~pixels of the pixel boundary and found that the best-fit phase curve parameters did not change significantly. In this paper, we have chosen to present the results from the untrimmed fit.

The presence of uncorrected systematics can bias our estimates of the best-fit eclipse depths, which subsequently affects our estimates of the dayside brightness temperature. To obtain a better estimate of the HAT-P-7b eclipse depths, we select a short 0.35-day (16,000 data points) segment of the phase curve observation surrounding each eclipse and fit the data to a simplified secondary eclipse light curve model using PLD to decorrelate the intrapixel sensitivity variations, as described in \citet{wong2}. The HAT-P-7b eclipse data with instrumental effects removed are shown in Figure~\ref{eclipses2} along with the best-fit secondary eclipse model light curves. In Table~\ref{tab:values2}, we report the eclipse parameters for HAT-P-7b  derived from our individual eclipse fits. We also tried fitting each WASP-19b eclipse individually and found that the depths derived from our individual eclipse fit are consistent with the depths computed from the global phase curve fit at better than the $0.2\sigma$ level. We choose to report the eclipse depths from the global WASP-19b phase curve fits in this paper.

For WASP-19b, the error-weighted average eclipse depths are $0.485\%\pm0.024\%$ and $0.584\%\pm 0.029\%$ for the 3.6 and 4.5~$\mu$m bandpasses, respectively. These are consistent with the depths reported in \citet{anderson} in their analysis of previous \textit{Spitzer} secondary eclipse observations  at better than the $1\sigma$ level ($0.483\%\pm0.025\%$ at 3.6~$\mu$m and $0.572\%\pm0.030\%$ at 4.5~$\mu$m). An analogous calculation for HAT-P-7b yields error-weighted average 3.6 and 4.5~$\mu$m eclipse depths of $0.156\%\pm0.009\%$ and $0.190\%\pm 0.006\%$, respectively. These values differ from the previously reported depths in \citet{christiansen} at the $3.0\sigma$ and $1.4\sigma$ levels, respectively ($0.098\%\pm0.017\%$ at 3.6~$\mu$m and $0.159\%\pm0.022\%$ at 4.5~$\mu$m). The \citet{christiansen} study analyzed secondary eclipse observations which were obtained in the full-array mode of \textit{Spitzer}/IRAC during the cryogenic mission and were analyzed using the standard (at the time) second-order polynomial decorrelation technique, which has since been shown to lead to biases in the estimated eclipse depths in a subset of cases \citep[e.g.,][]{diamondlowe, hansen, lanotte, pld}. We note that the previous \citet{anderson} analysis of WASP-19b \textit{Spitzer} secondary eclipses also utilized the polynomial decorrelation technique, but contrary to the case of HAT-P-7b, this approach did not produce a discernible bias in the estimated eclipse depths.

\begin{figure}[t]
	\begin{center}
		\includegraphics[trim=0 0 0 9cm,clip,width=9cm]{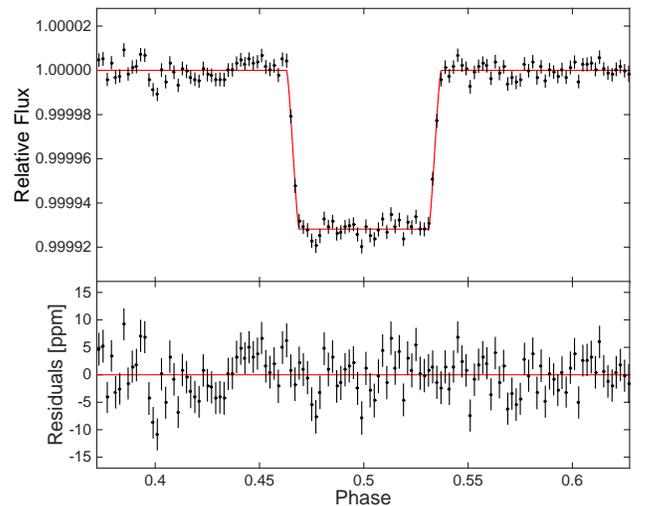}
	\end{center}
	\caption{Top: plot of the phase-folded HAT-P-7b {\it Kepler} secondary eclipse data, binned in intervals of 0.002 in phase (black dots), and the best-fit light curve solution (red line). Bottom: the corresponding residuals from the best-fit solution.}\label{kepler}
\end{figure}

HAT-P-7b resides in the {\it Kepler} field, making it one of the few transiting exoplanets for which the secondary eclipse depth has been measured in both the visible light and the infrared, and the only one so far with full orbital phase curves in both wavelength regimes. The planet was discovered prior to {\it Kepler}'s launch \citep{pal} and has the alternative designation Kepler-2b. It was observed by {\it Kepler} in short cadence mode with a one-minute exposure time throughout the entire {\it Kepler} mission \citep{gilliland}. In Figure~\ref{kepler}, we show the phase-folded and binned {\it Kepler} light curve of the secondary eclipse, which we have analyzed here. Our analysis is similar to that of the Kepler-13Ab secondary eclipse presented in \citet{shporer2014}, where only the eclipse depth and mid-eclipse time are free parameters in the model fitting, with the rest of the parameters constrained to their known values and uncertainties. The results for the two fitted parameters are $t_{E,\mathrm{Kep}} = 2455709.48109 \pm 0.00013$~(BJD$_{\mathrm{TDB}}$) and $d_{\mathrm{Kep}}=71.80\pm0.32$~ppm. This eclipse depth is consistent (within $1\sigma$) with some of the values published by others \citep{coughlin,vaneylen,morris,esteves2015}, while lying a few $\sigma$ away from other values reported in the literature \citep{borucki,jackson,esteves2013,angerhausen}. The reasons for the difference are not immediately apparent, but likely include the smaller amount of \textit{Kepler} data analyzed in the earlier studies and differences in detrending techniques and the handling of correlated noise.

\begin{table}[t!]
\centering
\begin{threeparttable}
\caption{Updated WASP-19b Planetary Parameters} \label{tab:newvalues1}

\renewcommand{\arraystretch}{1.2}

\begin{tabular}{ l m{0.001cm} c  }
\hline\hline
Parameter& & Value \\
  
 \hline 
\vspace{1mm}
$P$~(days) & & $0.788838989 \pm 0.000000040$ \\
$T_{c,0}$~(BJD$_{\mathrm{TDB}}$) & & $2455708.534626 \pm 0.000019$\\
$R_{p}/R_{*}$ & & $0.1409\pm 0.0013$ \\
$a/R_{*}$ & & $3.46\pm 0.08$\\
$i$~$(^{\circ})$ & & $78.78\pm 0.58$\\
$e$ & & $0.002^{+0.014}_{-0.002}$ \\
$\omega$~$(^{\circ})$& & $259^{+13}_{-170}$ \\
$M_{p}$~$(M_{\mathrm{Jup}})$ & & $1.069^{+0.038}_{-0.037}$ \\
$R_{p}$~$(R_{\mathrm{Jup}})$ & & $1.392\pm 0.040$\\
$\rho_{p}$~(g cm$^{-3}$) & & $0.492\pm 0.046$\\
$g_{p}$~(m s$^{-2}$) & & $13.68\pm 0.92$\\
$a$~(au) & & $0.01634\pm 0.00024$ \\
\hline

\end{tabular}
      \end{threeparttable}
\end{table}

\begin{table}[t!]
\centering
\begin{threeparttable}
\caption{Updated HAT-P-7b Planetary Parameters} \label{tab:newvalues2}

\renewcommand{\arraystretch}{1.2}

\begin{tabular}{ l m{0.001cm} c  }
\hline\hline
 Parameter& & Value \\
   
 \hline 
\vspace{1mm}
$P$~(days) & & $2.2047372\pm 0.0000011$ \\
$T_{c,0}$~(BJD$_{\mathrm{TDB}}$) & & $2454731.68039 \pm 0.00023$\\
$R_{p}/R_{*}$ & & $0.07809\pm 0.00068$ \\
$a/R_{*}$ & & $4.03\pm 0.16$\\
$i$~$(^{\circ})$ & & $82.2\pm 1.2$\\
$e$ & & $0.0016^{+0.0034}_{-0.0010}$ \\
$\omega$~$(^{\circ})$& & $165^{+93}_{-66}$ \\
$M_{p}$~$(M_{\mathrm{Jup}})$ & & $1.682^{+0.021}_{-0.020}$ \\
$R_{p}$~$(R_{\mathrm{Jup}})$ & & $1.491\pm 0.061$\\
$\rho_{p}$~(g cm$^{-3}$) & & $0.629\pm 0.078$\\
$g_{p}$~(m s$^{-2}$) & & $18.8\pm 1.6$\\
$a$~(au) & & $0.03676\pm 0.00019$ \\
\hline

\end{tabular}
      \end{threeparttable}
\end{table}

\begin{figure}[t]
\begin{center}
\includegraphics[width=9cm]{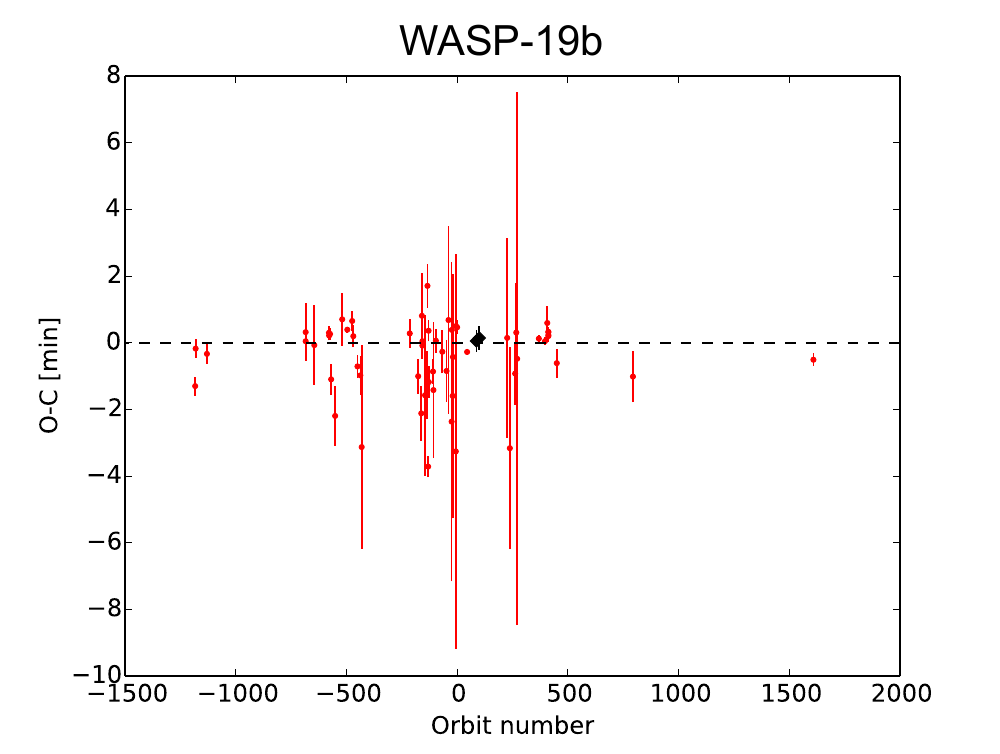}
\end{center}
\caption{Observed minus calculated WASP-19b transit times for all published observations \citep[red circles are previously published values;][]{hebb,anderson2010,dragomir,hellier,abe,bean,huitson,lendl,mancini,mandell,tregloanreed,sedaghati} based on the updated ephemeris derived in Section~\ref{sec:analysis}. The black diamonds denote the two transit times measured from our phase curve data.}\label{transitoc1}
\end{figure}

\begin{figure}[t]
\begin{center}
\includegraphics[width=9cm]{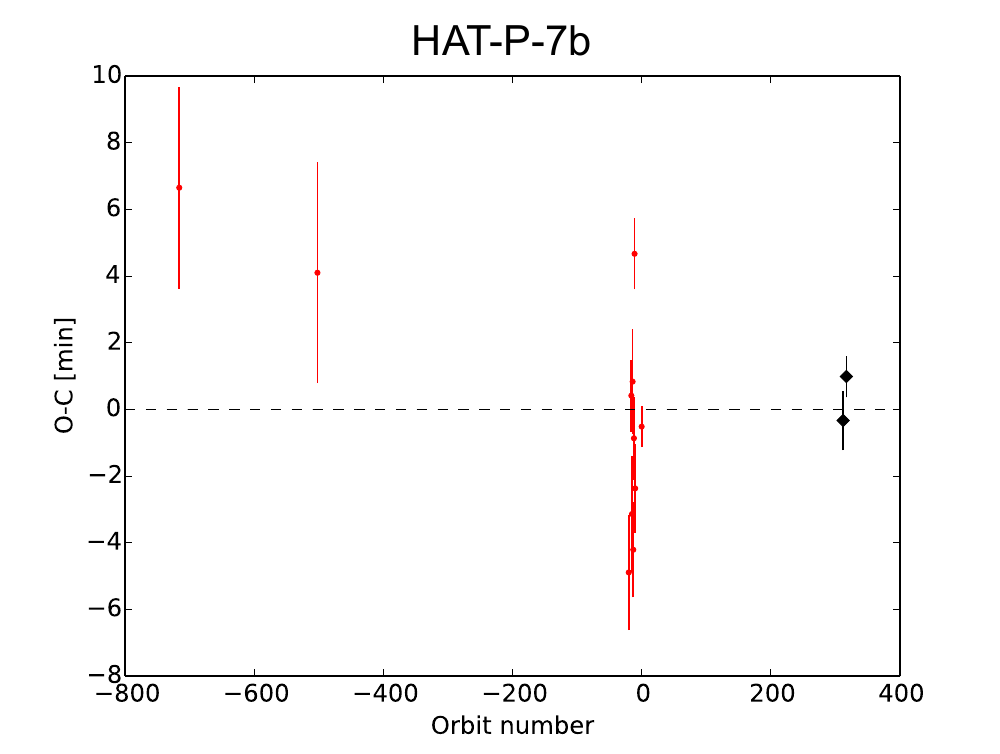}
\end{center}
\caption{Observed minus calculated HAT-P-7b transit times for all published observations \citep[red circles are previously published values;][]{pal,winn,christiansen} based on the updated ephemeris derived in Section~\ref{sec:analysis}. The black diamonds denote the two transit times measured from our phase curve data.}\label{transitoc2}
\end{figure}

\begin{figure}[t]
\begin{center}
\includegraphics[width=9cm]{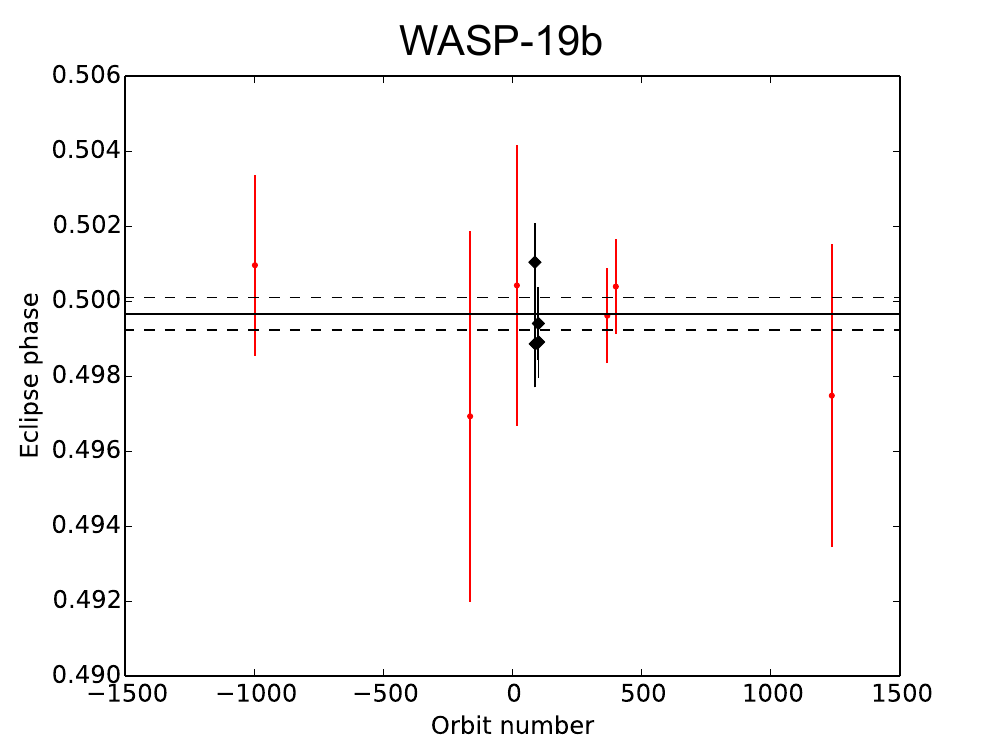}
\end{center}
\caption{Secondary eclipse phase of WASP-19b for all published observations \citep[red circles are previously published values from][]{gibson,burton,bean,mancini,zhou} based on the updated ephemeris calculated in Section~\ref{sec:analysis}. All eclipse times have been corrected for the light travel time across the system. The black diamonds denote the four secondary eclipse times measured from our phase curve data. The solid and dashed lines indicate the error-weighted mean phase value and corresponding $1\sigma$ confidence bounds, respectively.}\label{eclipseoc1}
\end{figure}

\begin{figure}[t]
\begin{center}
\includegraphics[width=9cm]{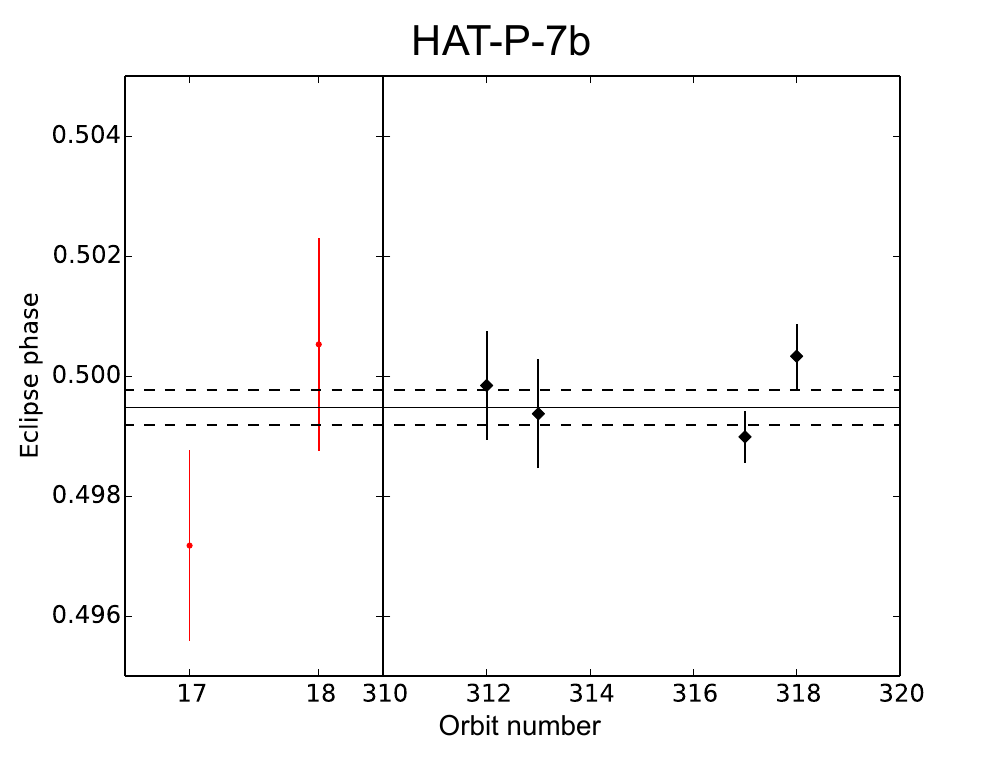}
\end{center}
\caption{Secondary eclipse phase of HAT-P-7b for all published observations \citep[red circles are previously published values from][]{christiansen} based on the updated ephemeris calculated in Section~\ref{sec:analysis}. All eclipse times have been corrected for the light travel time across the system. The black diamonds denote the four secondary eclipse times measured from our phase curve data. The solid and dashed lines indicate the error-weighted mean phase value and corresponding $1\sigma$ confidence bounds, respectively.}\label{eclipseoc2}
\end{figure}

We use standard methods \citep[e.g.,][]{wong2} to compute an updated ephemeris for each planet using the transit times calculated  from our phase curve observations and all previously published, independent, single-epoch transit times. In those cases where the timing standard (UTC vs. TDB) was not indicated, we assume UTC. The updated estimates of the orbital period $P$ and zeroth epoch mid-transit time $T_{c,0}$ are listed in Tables~\ref{tab:newvalues1} and \ref{tab:newvalues2} for WASP-19b and HAT-P-7b, respectively. The observed minus calculated transit times are shown in Figures~\ref{transitoc1} and \ref{transitoc2}. We also derive an independent estimate of the orbital period by fitting through all published secondary eclipse times and obtain $P=0.7888385 \pm 0.0000013$ (WASP-19b) and $P =2.2047436\pm 0.0000090$ (HAT-P-7b), which are consistent with the corresponding best-fit transit periods at better than the $1\sigma$ level. Using the updated transit ephemeris values, we compute the orbital phase of secondary eclipse for each eclipse (Figures~\ref{eclipseoc1} and \ref{eclipseoc2}) and derive error-weighted values of $0.49967\pm 0.00043$ for WASP-19b and $0.49948 \pm 0.00029$ for HAT-P-7b.

\begin{figure}[t!]
\begin{center}
\includegraphics[width=8.7cm]{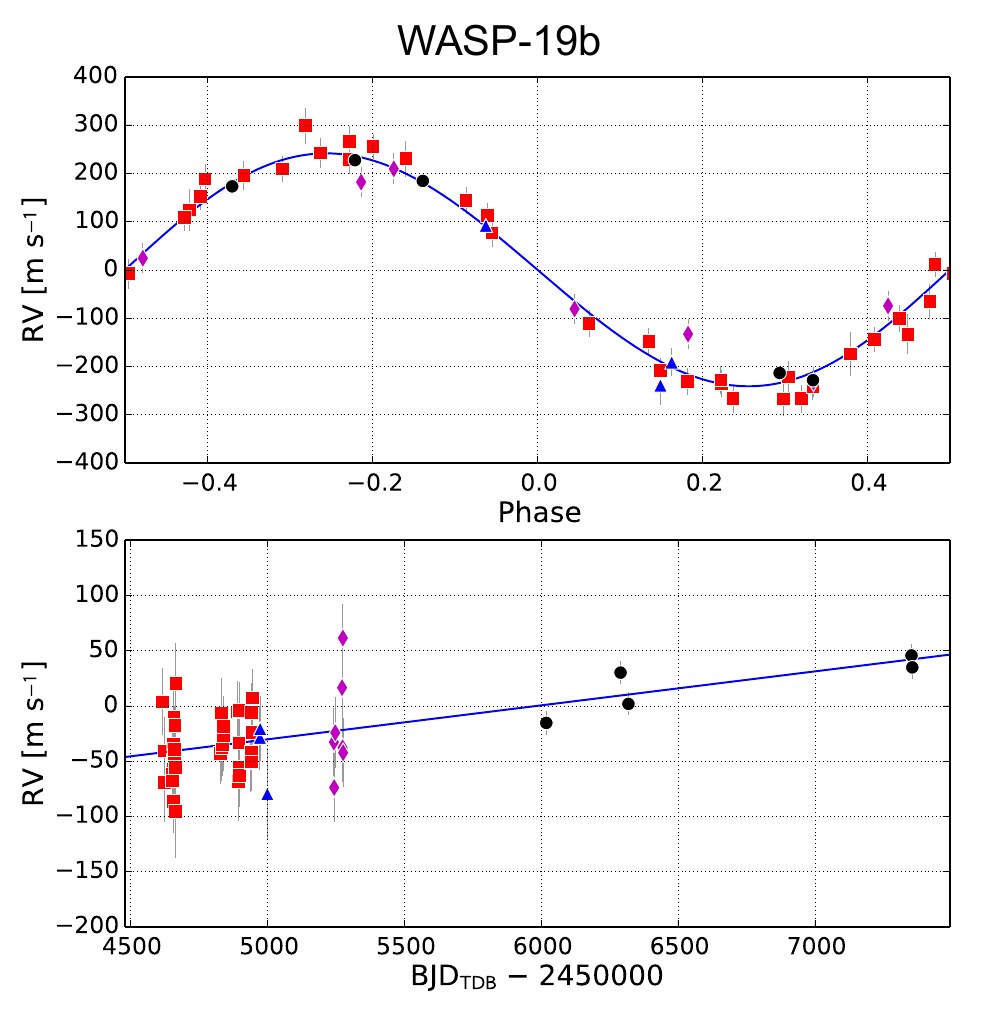}
\end{center}
\caption{Top panel: phased best-fit radial velocity curve of all published radial velocity measurements of WASP-19. The points from each data set are color-coded: black circles --- Keck/HIRES \citep{knutson2014}, red squares --- CORALIE \citep{hebb}, blue triangles --- CORALIE \citep{hellier}, magenta diamonds --- HARPS \citep{hellier}. Bottom panel: corresponding unphased residuals after the radial velocity solution for the transiting hot Jupiter is removed. We detect a linear trend in our RV fits at the $2.8\sigma$ level.} \label{rv1}
\end{figure}

\begin{figure}[t!]
\begin{center}
\includegraphics[width=8.7cm]{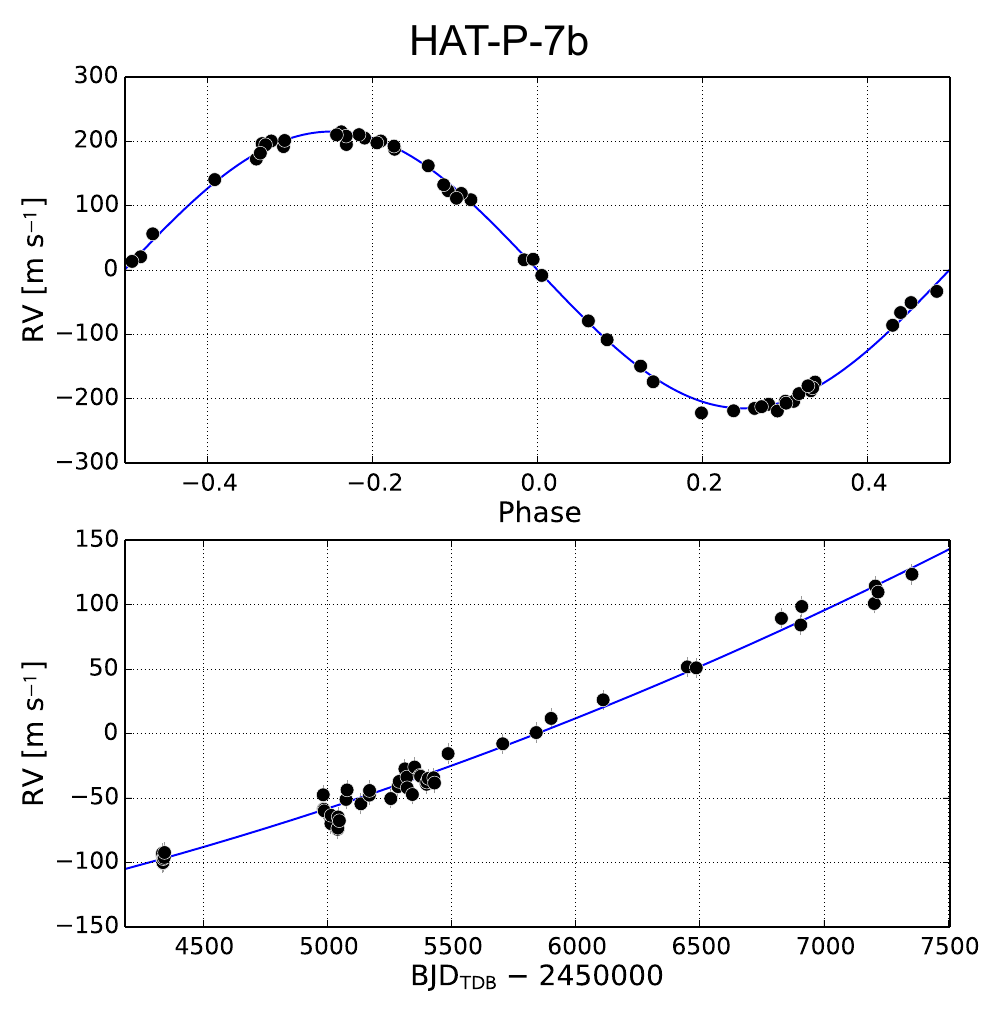}
\end{center}
\caption{Top panel: phased radial velocity curve of radial velocity measurements of HAT-P-7 published in \citet{knutson2014}, with six additional measurements obtained using the Keck/HIRES instrument. Bottom panel: corresponding unphased residuals after the radial velocity solution for the transiting hot Jupiter is removed. The robust detection of an acceleration in the radial velocity data suggests the presence of a brown dwarf or planetary companion.} \label{rv2}
\end{figure}

We carry out an analysis of the RV measurements reported in \citet{knutson2014}, using the updated transit ephemeris and secondary eclipse times derived from our phase curve fits as priors; for HAT-P-7b, we have included an additional six measurements obtained using the Keck/HIRES instrument since the publication of \citet{knutson2014}. Comprehensive results from our RV fits are shown in Tables~\ref{tab:rv1}$-$\ref{tab:rv2} and Figures~\ref{rv1}$-$\ref{rv2}. These fits provide new estimates of the orbital period ($P_{b}$), center of transit time ($T_{\mathrm{conj},b}$), orbital eccentricity ($e_{b}$), and argument of perihelion ($\omega_{b}$), as well as updated values for the orbital semi-major axis and planet mass, from which we derive new estimates of the planetary radius, density, and surface gravity. The full lists of updated planetary parameters are given in Tables~\ref{tab:newvalues1} and \ref{tab:newvalues2}. 

A potential linear trend is detected in our WASP-19b RV fit at the $2.8\sigma$ level; additional measurements of this system are needed to determine whether this trend is indicative of a stellar or planetary companion. For the HAT-P-7 system, we find a non-zero RV slope (significant at above the $10\sigma$ level), which indicates the presence of one or more companions in the system. While a direct imaging survey revealed that HAT-P-7 has a stellar companion \citep{ngo}, it is too small and distant to explain the measured RV trend, which suggests that a closer-in brown dwarf or planetary companion is likely responsible for the observed RV trend.

\begin{table}[t!]
\centering
\begin{threeparttable}
\caption{Comprehensive results from radial velocity fit for WASP-19b \\ with priors on transit ephemeris and eclipse times} \label{tab:rv1}

\renewcommand{\arraystretch}{1.0}
\begin{center}
\begin{tabular}{l  r r}
\hline\hline
Parameter & Value & Units \\
\hline
  $P_{b}$ & 0.788839164 $\pm 4.2\mathrm{E}-08$ & days\\
$T_{\mathrm{conj},b}$ & 2455777.1636 $\pm 2.3\mathrm{E}-05$ & \bjdtdb\\
$e_{b}$ & 0.002 $^{+0.014}_{-0.002}$ & ...\\
$\omega_{b}$ & 259 $^{+13}_{-170}$ & degrees\\
$K_{b}$ & 241.9 $^{+4.6}_{-4.7}$ & \ms\\
$\gamma_{1}$ & $-$35.4 $^{+7.9}_{-7.7}$ & \ms\\
$\gamma_{2}$ & 76 $\pm 12$ & \ms\\
$\gamma_{3}$ & 95 $^{+20}_{-21}$ & \ms\\
$\gamma_{4}$ & 112 $\pm 13$ & \ms\\
$\dot{\gamma}$ & 0.0338 $\pm 0.0091$ & \ms day$^{-1}$\\
$\ddot{\gamma}$ & $\equiv$ 0.0 $\pm 0.0$ & \ms day$^{-2}$\\
jitter & $\equiv$ 10.0 $\pm 0.0$ & \ms\\
\hline
\end{tabular}

\begin{tablenotes}
      \small
       \item {\bf Notes.} Radial velocity zero point offsets ($\gamma_{1-4}$) derived from four separate RV data sets: 1 -- Keck/HIRES \citep{knutson2014}, 2 -- CORALIE \citep{hebb}, 3 -- CORALIE \citep{hellier}, 4 -- HARPS \citep{hellier}. $\dot{\gamma}$ and $\ddot{\gamma}$ are the slope and curvature of the best-fit radial velocity acceleration, respectively.  $K_{b}$ is the radial velocity semi-amplitude.
       \end{tablenotes}
    \end{center}
    \end{threeparttable}
\end{table}
    
\begin{table}[t!]
\centering
\begin{threeparttable}
\caption{Comprehensive results from radial velocity fit for HAT-P-7b \\ with priors on transit ephemeris and eclipse times} \label{tab:rv2}

\renewcommand{\arraystretch}{1.0}
\begin{center}
\begin{tabular}{l  r r}
\hline\hline
Parameter & Value & Units \\
\hline
$P_{b}$ & 2.2047375 $\pm 1.2\mathrm{E}-06$ & days\\
$T_{\mathrm{conj},b}$ & 2454731.68038 $^{+0.00023}_{-0.00024}$ & \bjdtdb\\
$e_{b}$ & 0.0016 $^{+0.0034}_{-0.001}$ & ...\\
$\omega_{b}$ & 165 $^{+93}_{-66}$ & degrees\\
$K_{b}$ & 213.8 $^{+1.3}_{-1.4}$ & \ms\\
$\gamma$ & 20.3 $^{+1.5}_{-1.6}$ & \ms\\
$\dot{\gamma}$ & 0.0753 $^{+0.0014}_{-0.0013}$ & \ms day$^{-1}$\\
$\ddot{\gamma}$ & 6.9$\mathrm{E}-$06 $\pm 1.3\mathrm{E}-06$ & \ms day$^{-2}$\\
jitter & 7.13 $^{+0.86}_{-0.73}$ & \ms\\
\hline
\end{tabular}

\begin{tablenotes}
      \small
       \item {\bf Notes.} Radial velocity zero point offset ($\gamma$) derived from the Keck/HIRES RV data set analyzed in \citep{knutson2014}, with six additional RV measurements obtained by the same instrument. See the text and Table~\ref{tab:rv1} for description of other variables.
\end{tablenotes}
    \end{center}
    \end{threeparttable}
\end{table}

\section{Discussion}\label{sec:disc}
\subsection{Phase curves}

\begin{figure*}[t]
\begin{center}
\includegraphics[width=18cm]{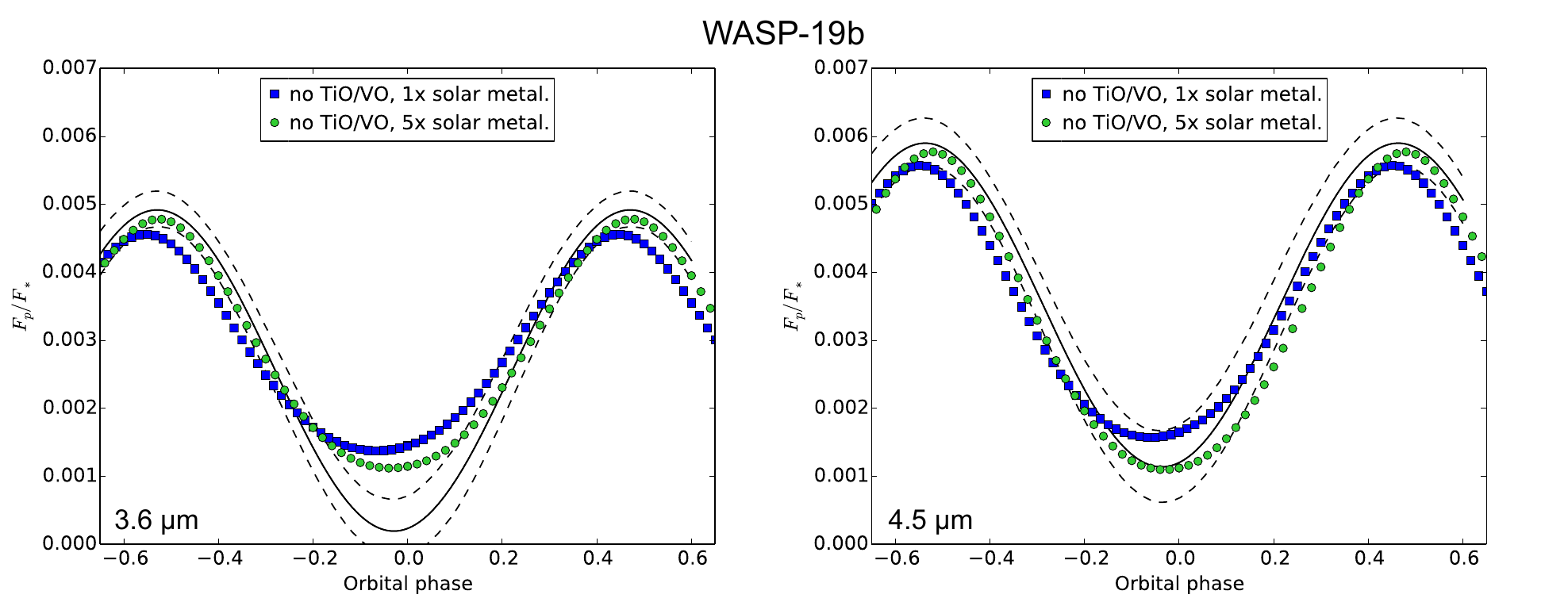}
\end{center}
\caption{Left: the best-fit WASP-19b 3.6~$\mu$m relative planet brightness  phase curve and $1\sigma$ brightness bounds (solid and dotted lines). The predicted phase curves from the no TiO/VO SPARC models with $1\times$ solar and $5\times$ solar atmospheric metallicity are plotted with blue squares and green circles, respectively. Right: analogous plot for the 4.5~$\mu$m bandpass.} \label{lightcurve1}
\end{figure*}

\begin{figure*}[t]
\begin{center}
\includegraphics[width=18cm]{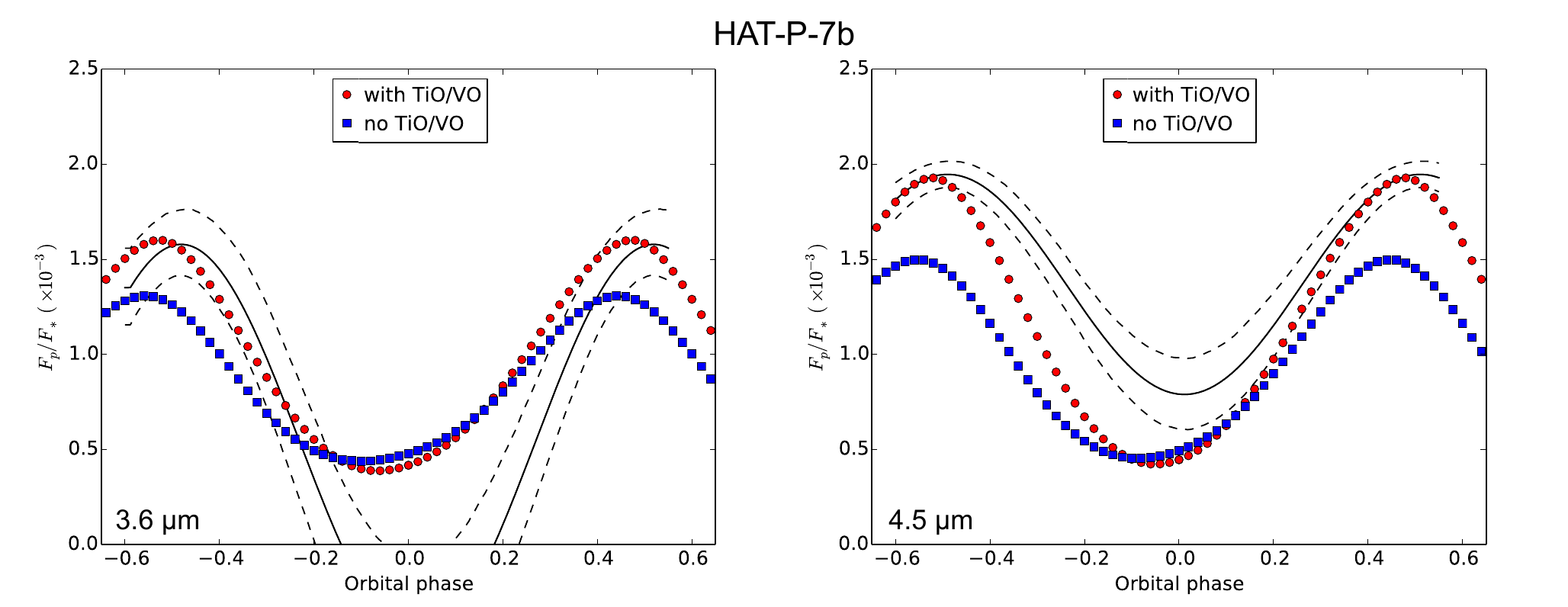}
\end{center}
\caption{Left: the best-fit HAT-P-7b 3.6~$\mu$m relative planet brightness  ratio phase curve and $1\sigma$ brightness bounds (solid and dotted lines). The predicted phase curves from the SPARC models with and without TiO/VO are plotted with red circles and blue squares, respectively. Both models assume solar metallicity. Right: analogous plot for the 4.5~$\mu$m bandpass.} \label{lightcurve2}
\end{figure*}

\begin{table}[t!]
\small
\centering
\begin{threeparttable}
\caption{WASP-19b Phase Curve Comparison} \label{tab:comparison1}

\renewcommand{\arraystretch}{1.0}
\begin{center}
\begin{tabular}{l  l l}
\hline\hline
Source & 3.6$\mu$m & 4.5$\mu$m \\
\hline

\sidehead{\textit{Maximum flux ratio [$\%$]}}Measured & $0.492^{+0.028}_{-0.024}$ & $0.590^{+0.037}_{-0.034}$\\
$1\times$ solar metal. model\textsuperscript{$^{\dag}$} & $0.456$ & $0.557$\\
$5\times$ solar metal. model\textsuperscript{$^{\dag}$} & $0.479$ & $0.577$ \\
\\
\sidehead{\textit{Minimum flux ratio [$\%$]}}Measured & $0.020^{+0.048}_{-0.038}$ & $0.114^{+0.053}_{-0.052}$\\
$1\times$ solar metal. model & $0.138$ & $0.157$\\
$5\times$ solar metal. model & $0.112$ & $0.110$ \\
\\
\sidehead{\textit{Phase curve amplitude [$\%$]}\textsuperscript{a}}Measured & $0.472^{+0.033}_{-0.036}$ & $0.476^{+0.046}_{-0.040}$\\
$1\times$ solar metal. model &$0.318$ & $0.400$\\
$5\times$ solar metal. model & $0.367$ & $0.467$ \\
\\
\sidehead{\textit{Maximum flux offset [h]}\textsuperscript{b}}Measured & $-0.55\pm0.21$ &  $-0.68\pm0.19$ \\
$1\times$ solar metal. model & $-0.95$ & $-0.95$\\
$5\times$ solar metal. model & $-0.38$ & $-0.38$ \\
\\
\sidehead{\textit{Minimum flux offset [h]}\textsuperscript{b}}Measured & $-0.55^{+0.24}_{-0.35}$ &  $-0.70^{+0.24}_{-0.30}$\\
$1\times$ solar metal. model & $-1.26$ & $-1.26$\\
$5\times$ solar metal. model & $-0.76$ & $-0.76$ \\
 \hline
\end{tabular}

\begin{tablenotes}
      \small
       \item {\bf Notes.} 
       \item \textsuperscript{$^{\dag}$}{Both models here have no TiO or VO.}
	\item \textsuperscript{a}{Difference between maximum and minimum flux ratios.}
	\item \textsuperscript{b}{The maximum and minimum flux offsets are measured relative to the center of secondary eclipse time and center of transit time, respectively. Negative time offsets indicate an eastward shift in the location of the hot or cold region in the planet's atmosphere.}
\end{tablenotes}
    \end{center}
    \end{threeparttable}
\end{table}

\begin{table}[t!]
\small
\centering
\begin{threeparttable}
\caption{HAT-P-7b Phase Curve Comparison} \label{tab:comparison2}

\renewcommand{\arraystretch}{1.0}
\begin{center}
\begin{tabular}{l  l l}
\hline\hline
Source & 3.6$\mu$m & 4.5$\mu$m \\
\hline
\sidehead{\textit{Maximum flux ratio [$\%$]}}Measured & $0.158^{+0.019}_{-0.017}$ & $0.195\pm 0.007$\\
With TiO/VO model\textsuperscript{a} & $0.160$ & $0.193$\\
No TiO/VO model\textsuperscript{a} & $0.131$ & $0.150$\\
\\
\sidehead{\textit{Minimum flux ratio [$\%$]}}Measured & $<0.030$\textsuperscript{b} & $0.079\pm0.019$\\
With TiO/VO model & $0.039$ & $0.042$\\
No TiO/VO model & $0.044$ & $0.045$\\
\\
\sidehead{\textit{Phase curve amplitude [$\%$]}\textsuperscript{c}}Measured & $>0.128$\textsuperscript{b} & $0.116^{+0.018}_{-0.019}$\\
With TiO/VO model & $0.121$ & $0.151$\\
No TiO/VO model &$0.089$ & $0.105$\\
\\
\sidehead{\textit{Maximum flux offset [h]}\textsuperscript{d}}Measured & $1.0\pm 1.1$ & $0.6\pm 1.1$\\
With TiO/VO model & $-1.1$ & $-1.1$\\
No TiO/VO model & $-3.2$ & $-3.2$\\
\\
\sidehead{\textit{Minimum flux offset [h]}\textsuperscript{d}}Measured & $1.1^{+1.5}_{-1.6}$ &  $0.6^{+1.5}_{-1.7}$\\
With TiO/VO model & $-3.2$ & $-2.1$\\
No TiO/VO model & $-5.3$ & $-4.2$\\
 \hline
\end{tabular}

\begin{tablenotes}
      \small
       \item {\bf Notes.} 
       \item\textsuperscript{a}{Both models for HAT-P-7b assume solar metallicity.}
 	\item\textsuperscript{b}{Based on a $2\sigma$ upper limit of the minimum flux ratio.}
	\item \textsuperscript{c}{Difference between maximum and minimum flux ratios.}
	\item \textsuperscript{d}{The maximum and minimum flux offsets are measured relative to the center of secondary eclipse time and center of transit time, respectively. Negative time offsets indicate an eastward shift in the location of the hot or cold region in the planet's atmosphere.}
\end{tablenotes}
    \end{center}
    \end{threeparttable}
\end{table}

After subtracting and dividing out the brightness of the star alone, i.e., the flux measured during secondary eclipse, we obtain the phase variation of the relative planet--star flux ratio (Figures~\ref{lightcurve1} and \ref{lightcurve2}). Tables~\ref{tab:comparison1} and \ref{tab:comparison2} list the important quantitative characteristics of the WASP-19b and HAT-P-7b phase curves, respectively. By examining the properties of these phase curves, we obtain a basic picture of the temperature distribution across the planets' surfaces, from which we can begin to infer aspects of the planets' energy balance, heat transport, and day--night recirculation efficiency \citep[e.g.,][]{cowanagol}. WASP-19b and HAT-P-7b are particularly well-suited to a comparative phase curve study as they have similar levels of irradiation and surface gravity, both of which are predicted to be important factors in shaping the atmospheric circulation patterns. Despite these similarities, their orbital periods and corresponding synchronous rotation periods differ by a factor of three, making these planets an interesting test case for the importance of rotation as a driver of atmospheric dynamics.

Theoretical studies \citep[e.g.,][]{showmanguillot,cooper} and numerical models \citep[including three-dimensional general circulation models; e.g.,][]{showman2009,showman2015,lewis2010,rm2010,rm2012,rm2013,heng2011a,heng2011b,perna2,dobbs,kataria} both predict that the hot spots in the dayside atmospheres of these planets will be shifted eastward relative to the substellar point due to the presence of a superrotating, eastward-flowing equatorial jet. Such an eastward-shifted hot spot is manifested in the phase curve as a planetary brightness maximum that occurs prior to the center of eclipse time (i.e., a negative phase offset). Since the equatorial jet transfers hot gas from the dayside over to the unirradiated nightside, the presence of a significant hotspot offset also typically entails relatively efficient day--night energy transport.

For WASP-19b, the planetary brightness in both \textit{Spitzer} bandpasses peaks prior to the time of secondary eclipse, which implies the presence of an eastward-shifted hot spot on the dayside hemisphere and relatively efficient day--night heat transport. Similarly, the minimum flux offsets computed for WASP-19b are also negative, indicating that the coldest region of the atmosphere is shifted eastward from the anti-stellar point. The maximum and minimum flux offsets in the HAT-P-7b phase curves are both consistent with zero, suggesting that the hotspot is located near the substellar point and that the day--night recirculation is relatively inefficient. The best-fit planet--star flux ratio for HAT-P-7b at 3.6~$\mu$m dips below zero on the nightside and is consistent with zero at the $1-2\sigma$ level; this behavior was also reported in our previous analysis of the WASP-14b multiband \textit{Spitzer} phase curves \citep{wong2}. Henceforth, we place a $2\sigma$ upper limit on the measured HAT-P-7b planetary brightness  on the nightside.

The HAT-P-7b {\it Kepler} phase curve was analyzed by several authors \citep[e.g.,][]{borucki,welsh,jackson,vaneylen,esteves2013,angerhausen,esteves2015,faigler2015}. It is interesting to compare the results obtained here from the analysis of the \textit{Spitzer} infrared phase curves to those derived from the {\it Kepler} phase curve. While infrared phase curves are dominated by processes taking place in the planetary atmosphere, namely, thermal emission and possibly also reflected light, the visible-light phase curves contain contributions from additional processes originating from the gravitational interaction between the planet and the star, such as distortions of the host star due to tidal interactions and the beaming effect \citep[i.e., Doppler boosting; e.g.,][]{loeb,zucker}. Therefore, gaining insight about the planetary atmosphere from the analysis of \textit{Kepler} phase curves requires that we account for all of the aforementioned processes. As a result, the phase variations induced by the planetary atmosphere alone cannot typically be viewed directly \citep[but see][for a few cases where this is possible]{shporerhu}.

\citet{esteves2015} identified an eastward atmospheric phase shift of $6.97\pm 0.30$~deg in the {\it Kepler} data, while \citet{faigler2015} measured it to be $8.0 \pm2.0$~deg and $5.4 \pm 1.5$~deg for the two models they used. Although small in value, those results are statistically significant and consistent with each other, suggesting that the visible-light phase shift may be caused by thermal emission from an eastward-shifted hot spot. This is consistent with the relatively high temperature of this exoplanet's dayside atmosphere \citep{esteves2015}.

We note that the best-fit infrared maximum flux offsets from our HAT-P-7b \textit{Spitzer} phase curve analysis are in the opposite (i.e., westward) direction (see Table~\ref{tab:comparison2}), but the uncertainties are $\gtrsim4$ times larger than those reported above for the visible-light \textit{Kepler} phase shift, which makes all of the infrared phase shifts consistent with zero. In addition, the measured infrared phase shifts are all within $1-€"2\sigma$ of the visible-light phase shifts reported in the literature. Furthermore, the \textit{Kepler} bandpass typically probes deeper pressures than the \textit{Spitzer} bandpasses where the radiative timescale is longer, and so it is possible that the visible-light phase curve shows an eastward hotspot offset, while the infrared phase curves are consistent with no phase offset. Although  a measurement of a phase shift in the infrared could, in principle, have shown that the eastward phase shift seen in the visible phase curves is due to thermal emission from an offset hotspot, the infrared data are not sensitive enough confirm or refute this.

\subsection{Brightness temperature}\label{subsec:temps}

Assuming zero albedo, the predicted dayside equilibrium temperature of WASP-19b is 2520~K if the incident energy from the star is re-radiated from the dayside only and 2120~K if the planet re-radiates uniformly over its entire surface. A higher theoretical upper limit for the dayside temperature is obtained in the case of zero albedo and zero recirculation when we assume that each region on the dayside is a blackbody in local equilibrium with the incident stellar flux: 2710~K \citep{cowanagol2011}. From the measured secondary eclipse depths and interpolated PHOENIX spectra for the host star  \citep{husser}, we obtain best-fit dayside brightness temperatures of $2384^{+41}_{-57}$~K at 3.6~$\mu$m and $2357\pm64$~K at 4.5~$\mu$m.  We also find that the dayside planetary brightnesses in both bands are well-fit by a single blackbody with an effective temperature of $T_{\rm eff} = 2372^{+59}_{-60}$~K.  The brightness temperatures from our blackbody fits lie intermediate to the theoretical limits given above, suggesting that WASP-19b has relatively efficient day--night recirculation at the pressures probed by the \textit{Spitzer} bandpasses, which is consistent with the previously noted detection of a significant hotspot offset in the dayside atmosphere.
 
An analogous calculation for HAT-P-7b yields brightness temperatures of $2632 \pm 77$ and $2682\pm 49$ at 3.6 and 4.5~$\mu$m, respectively. We also find that the dayside planetary brightnesses in both bands are well-fit by a single blackbody with an effective temperature of $T_{\rm eff} = 2667\pm 57$~K. The predicted equilibrium temperatures assuming zero albedo are 2700 and 2270~K in the cases of zero and complete day--night recirculation, respectively. If we assume local equilibrium in the zero-albedo, no-recirculation limit, then we obtain a predicted dayside temperature of 2900~K. Our measured effective dayside temperature for HAT-P-7b is closer to the theoretical upper limit than for WASP-19b, which suggests that HAT-P-7b has less efficient day--night recirculation than WASP-19b, which is in agreement with the non-detection of a significant phase offset in the HAT-P-7b phase curves.
 
\begin{figure}[t]
\begin{center}
\includegraphics[width=9cm]{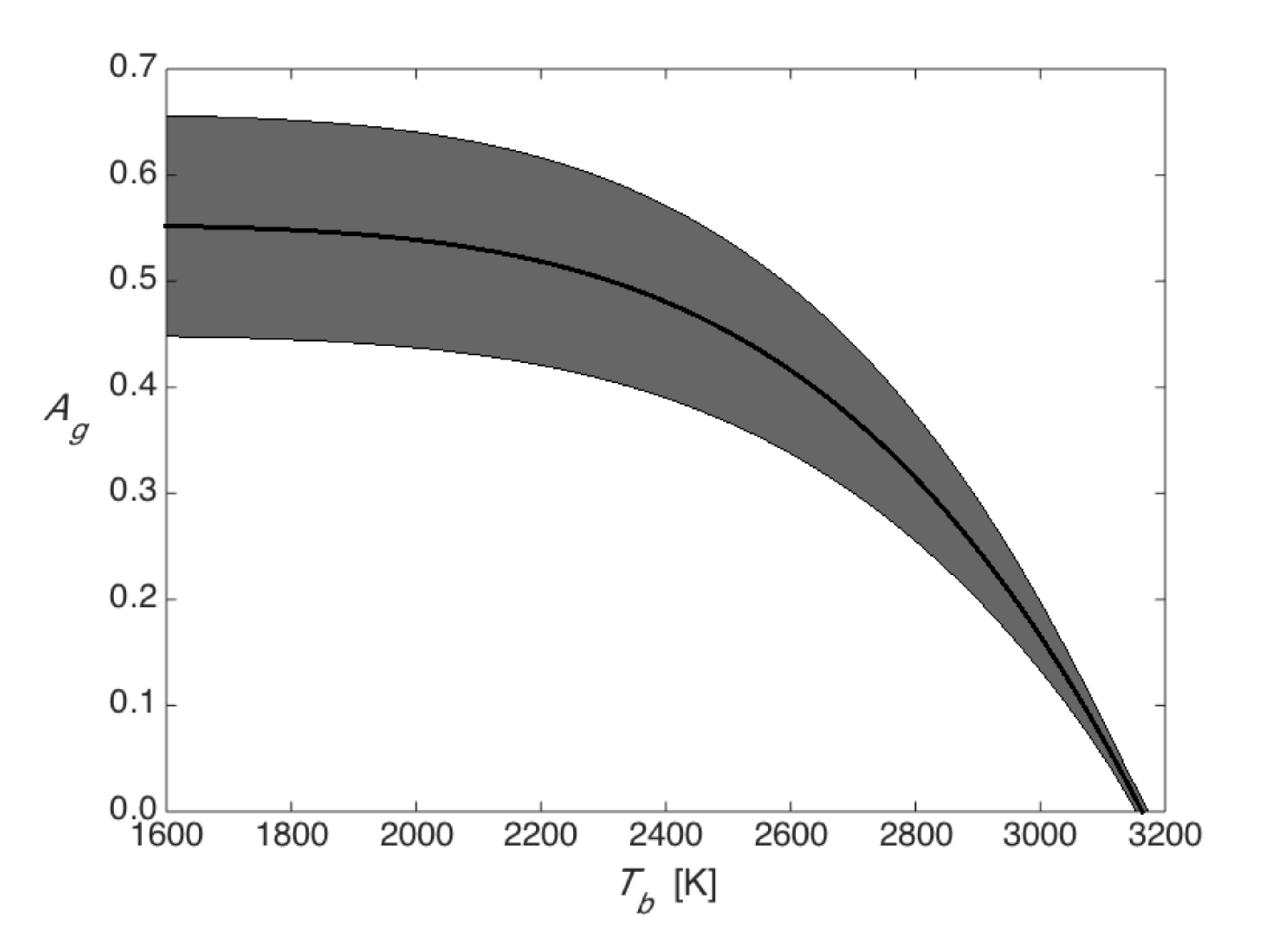}
\end{center}
\caption{Plot showing the relationship between the brightness temperature of HAT-P-7b derived from the optical \textit{Kepler} eclipse depth and the assumed geometric albedo.}\label{keplerkepler}
\end{figure}

We use the \textit{Kepler} eclipse depth ($d_{\mathrm{Kep}}=71.80\pm0.32$~ppm) to derive independent estimates of the dayside brightness temperature of HAT-P-7b, following the methods described in \citet{shporer2014}. Since the optical eclipse depth includes the contribution from the planet's emission as well as reflected starlight, the brightness temperature estimate depends on the assumed geometric albedo. An upper bound of $3160\pm10$~K is obtained when assuming zero albedo. On the cold end, the contribution of thermal emission to the eclipse depth becomes negligible as the geometric albedo asymptotically approaches 0.55. At brightness temperatures below 2320~K, thermal emission contributes less than 10\% to the eclipse depth, and the geometric albedo becomes larger than 0.50; below 1815~K, the contribution is less than 1\%, placing the geometric albedo at the 0.55 limit. The dayside brightness temperatures derived above from the infrared \textit{Spitzer} eclipse depths lie between these bounds. Figure~\ref{keplerkepler} shows the dependence of the \textit{Kepler}-derived brightness temperature on the assumed geometric albedo.

We can also use the derived planetary fluxes at the time of transit to estimate the brightness temperature of the nightside atmospheres. For WASP-19b, the best-fit nightside brightness temperatures at 3.6 and 4.5~$\mu$m are 890~K, with a $1\sigma$ upper limits of 1170~K and $1130^{+240}_{-130}$~K, respectively; the lower limit on the planetary flux at 3.6~$\mu$m dips below zero. The nightside planetary fluxes are also consistent with a single blackbody with an effective temperature of $1090^{+190}_{-250}$~K. For HAT-P-7b, the nightside planetary brightness drops below detectable levels in the 3.6~$\mu$m bandpass, and so we use the $2\sigma$ upper limit to calculate a corresponding brightness temperature of 1360~K. At 4.5~$\mu$m, the brightness temperature is $1710\pm 180$~K.

\subsection{Comparison with theoretical models}

As in our previous study of WASP-14b \citep{wong2}, we compare the results from our phase curve analysis with  light curves generated from the 3D Substellar and Planetary Atmospheric Radiation and Circulation (SPARC) GCM \citep{showman2009} and theoretical spectra produced by the 1D radiative transfer atmospheric model described in \citet{burrows}. The details of the models are summarized in \citet{wong2} and references therein.

\subsubsection{Three-dimensional GCM light curves}

The models we consider in this paper assume local thermochemical equilibrium and solar composition both with and without the incorporation of the strong optical absorbers TiO and VO (hereafter, ``with TiO/VO'' and ``no TiO/VO'', respectively), which are responsible for developing a dayside temperature inversion \citep{fortney}. In the case of WASP-19b, we also explore a model with enhanced metallicity.

\begin{figure}[t]
\begin{center}
\includegraphics[width=9cm]{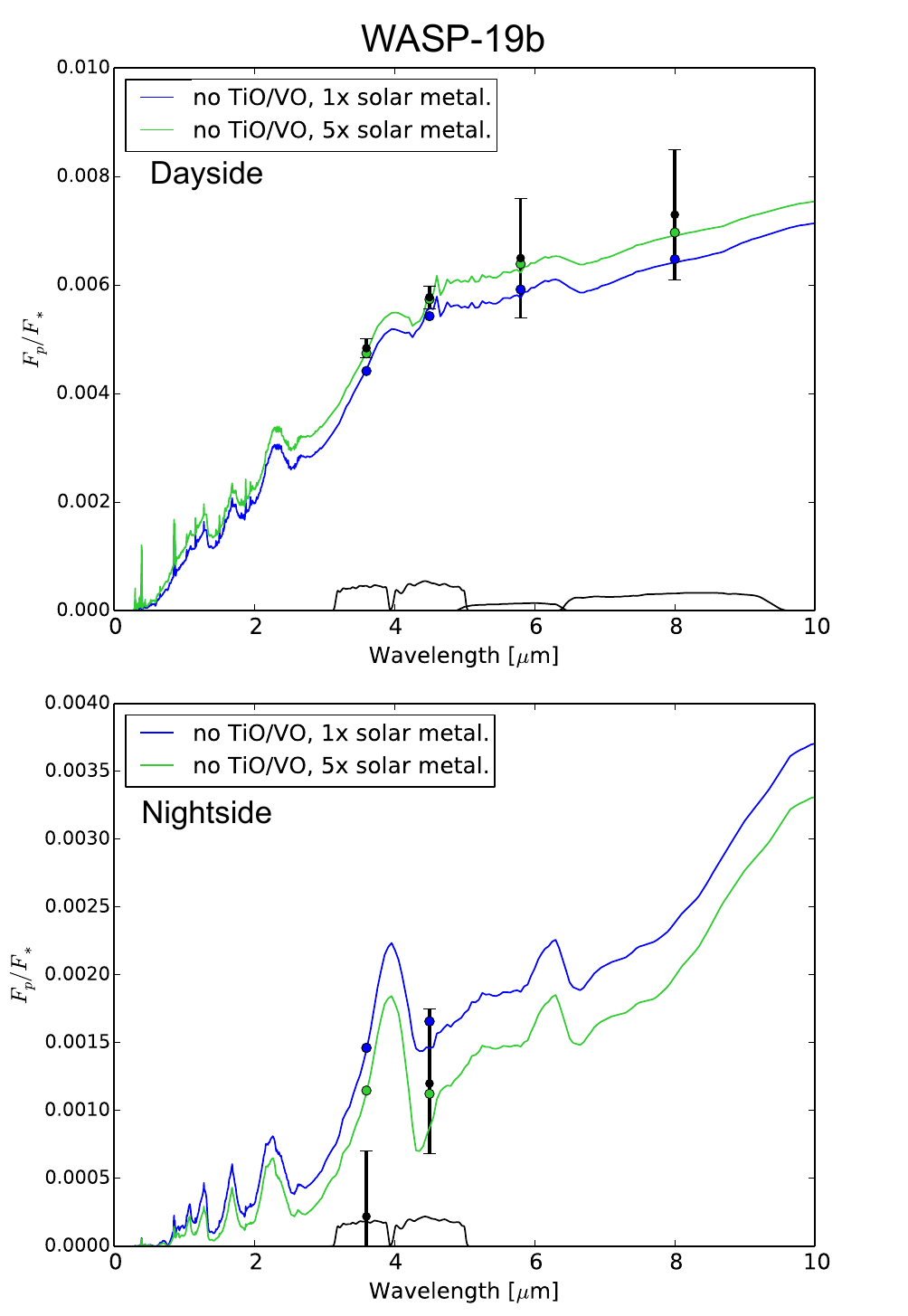}
\end{center}
\caption{Comparison of the error-weighted average measured dayside and nightside relative planetary brightnesses (filled black circles) for WASP-19b with SPARC model emission spectra at the time of secondary eclipse and the time of transit. The previously published 5.8 and 8.0~$\mu$m eclipse depths reported in \citet{anderson} are also included. The solid blue and green lines indicate the model-generated spectra for models with no TiO/VO, assuming $1\times$ solar and $5\times$ solar metallicities, respectively. The corresponding band-averaged fluxes are overplotted as filled points of the same color. The black lines at the bottom represent the photometric band transmission profiles in arbitrary units. All of the measured dayside planetary fluxes are well-matched by both SPARC model spectra. Meanwhile, the measured nightside planetary fluxes are most consistent with the $5\times$ solar metallicty model, with the 3.6~$\mu$m emission being overestimated by both models.} \label{jonathan1}
\end{figure}

\begin{figure}[t]
\begin{center}
\includegraphics[width=9cm]{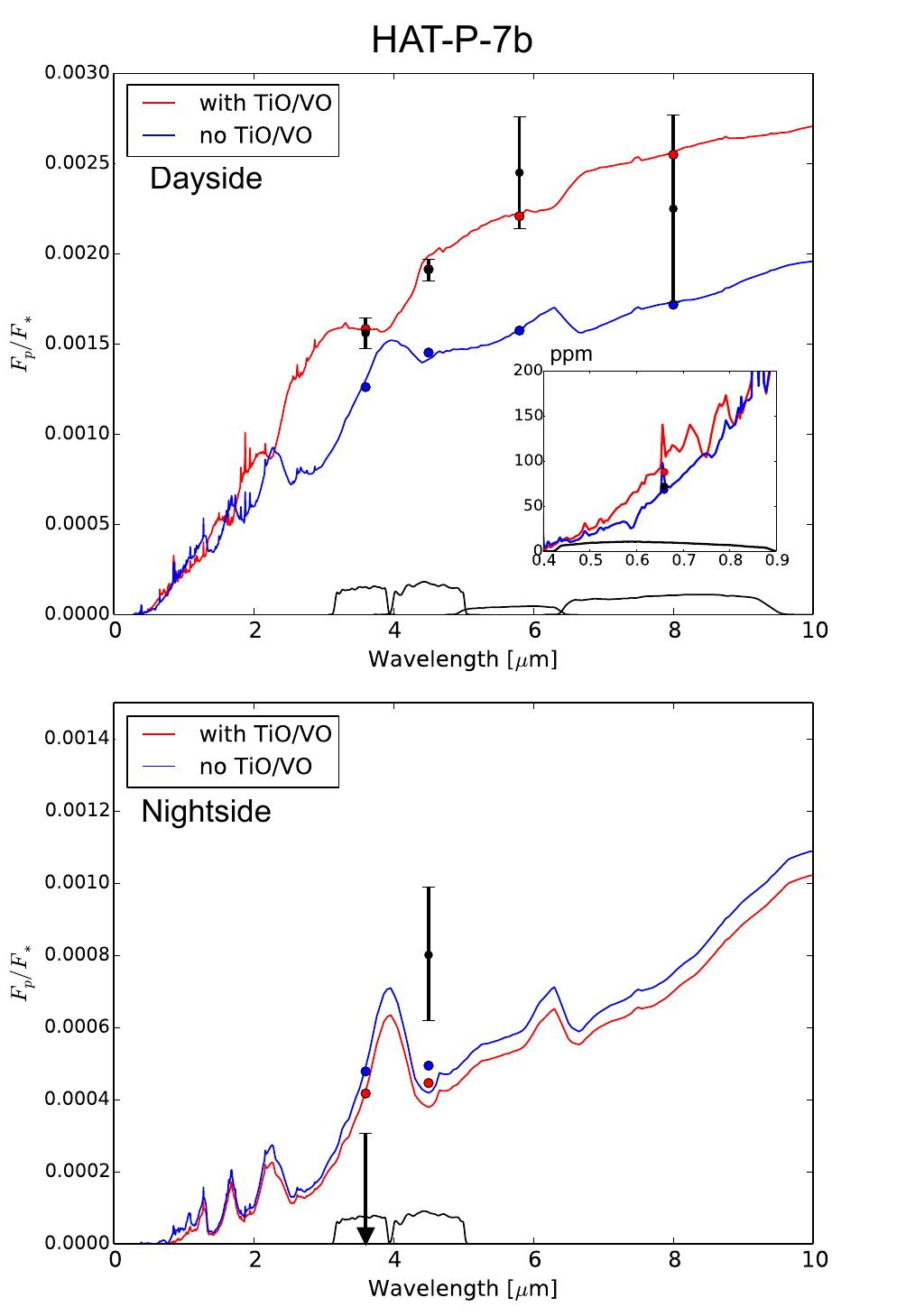}
\end{center}
\caption{Comparison of the measured dayside and nightside relative planetary brightnesses (filled black circles) for HAT-P-7b with SPARC model emission spectra. At 3.6~$\mu$m, the $2\sigma$ upper limit on the planetary brightness is shown. The previously published 5.8 and 8.0~$\mu$m eclipse depths reported in \citet{christiansen} are also included. The model-generated spectra for the with TiO/VO and no TiO/VO models are shown by the red and blue lines, respectively. Both models assume solar metallicity. The inset in the dayside comparison plot shows the measured eclipse depth in the \textit{Kepler} bandpass. The corresponding band-averaged fluxes are overplotted as filled points of the same color. The black lines at the bottom represent the photometric band transmission profiles in arbitrary units. The measured dayside planetary fluxes in infrared wavelengths are reasonably well-matched by the model spectrum with TiO/VO, while the nightside fluxes are not consistent with either model. The measured planetary brightness in the optical lies intermediate between the two models.} \label{jonathan2}
\end{figure}

\begin{figure}[t]
\begin{center}
\includegraphics[width=9cm]{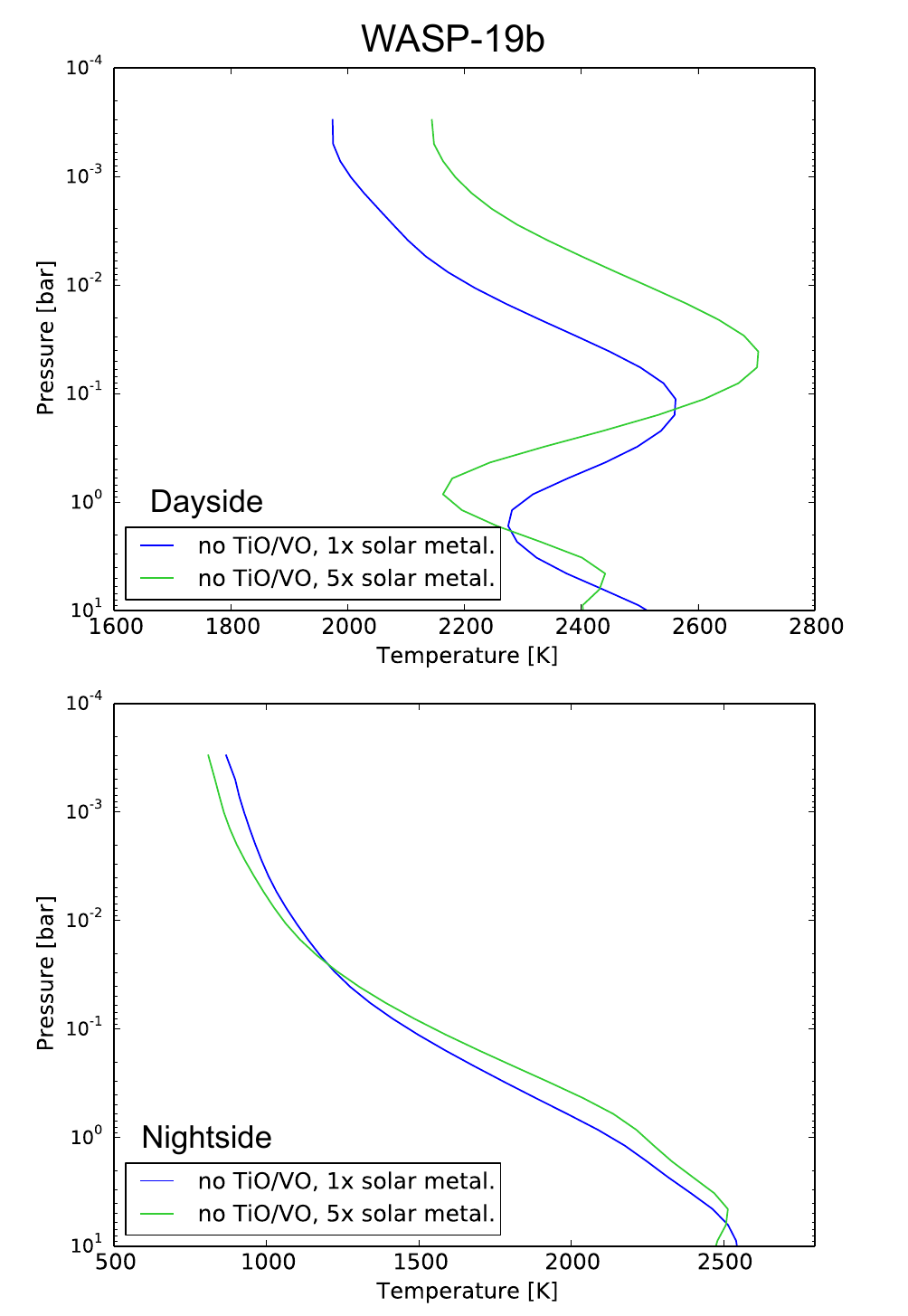}
\end{center}
\caption{SPARC model-generated temperature--pressure profiles for WASP-19b, corresponding to the emission spectra in Figure~\ref{jonathan1}.}\label{tp1}
\end{figure}

\begin{figure}[t]
\begin{center}
\includegraphics[width=9cm]{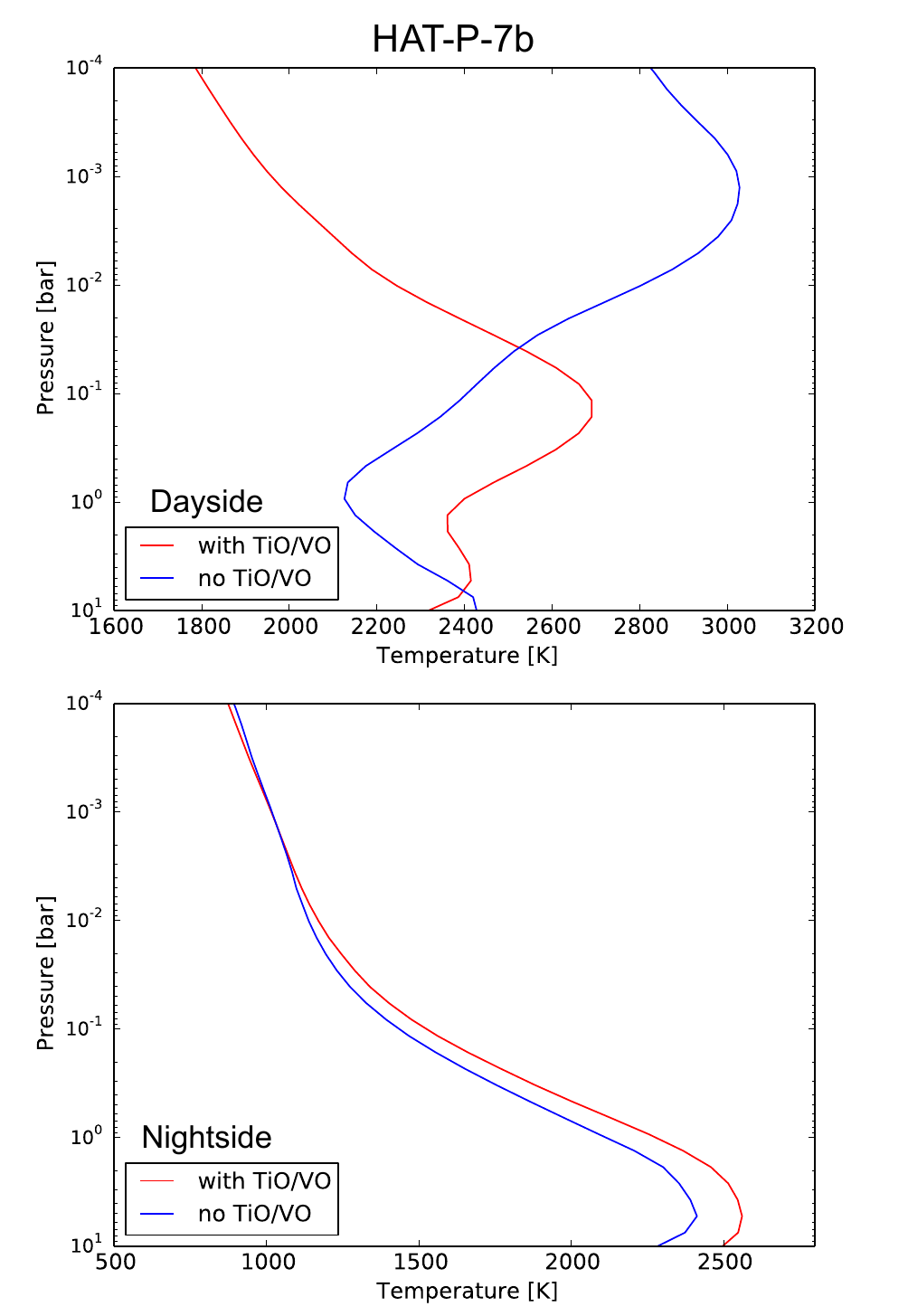}
\end{center}
\caption{SPARC model-generated temperature--pressure profiles for HAT-P-7b, corresponding to the emission spectra in Figure~\ref{jonathan2}.}\label{tp2}
\end{figure}

Figures~\ref{lightcurve1} and \ref{lightcurve2} compare the measured phase curves we obtained from fitting the \textit{Spitzer} data in each bandpass with the band-averaged theoretical light curves generated from the SPARC model using the methods of \citet{fortney2006}. The model-predicted maximum and minimum flux ratios and time offsets are listed in Tables~\ref{tab:comparison1} and \ref{tab:comparison2} for comparison with the corresponding values derived from our phase curve fits. Model-generated dayside and nightside spectra are shown in Figures~\ref{jonathan1} and \ref{jonathan2}, with the measured flux ratios overplotted. We have included the measured eclipse depth in the \textit{Kepler} bandpass for HAT-P-7b. The corresponding model-generated temperature--pressure profiles are shown in Figures~\ref{tp1} and \ref{tp2}.

For WASP-19b, we average the eclipse depths computed in the present work with previously published values from \citet{anderson} (consistent with our values at better than the $1\sigma$ level; Section~\ref{sec:analysis}) to arrive at $0.484\%\pm 0.017\%$  and $0.578\%\pm 0.021\%$ in the 3.6 and 4.5~$\mu$m bandpasses, respectively. We also include the measured 5.8 and 8.0~$\mu$m eclipse depths ($0.65\% \pm 0.11\%$ and $0.73\% \pm 0.12\%$, respectively) from \citet{anderson}. In light of the secondary eclipse analysis in \citet{anderson}, which using one-dimensional atmospheric models found that the dayside planetary emission is inconsistent with the presence of a temperature inversion, we only compare our WASP-19b data with the no TiO/VO SPARC model. For HAT-P-7b, we include the measured 5.8 and 8.0~$\mu$m eclipse depths ($0.245\% \pm 0.031\%$ and $0.225\% \pm 0.052\%$, respectively) from \citet{christiansen}, but do not average our 3.6 and 4.5~$\mu$m eclipse depths ($0.156\% \pm 0.009\%$ and $0.190\% \pm 0.006\%$, respectively) with their values, which differ significantly from ours (see Section~\ref{sec:analysis}). 

From Figure~\ref{jonathan1}, we see that the measured dayside planetary emission of WASP-19b is in good agreement with the theoretical spectrum generated from the solar metallicity no TiO/VO model in all bandpasses except 3.6~$\mu$m, where the model somewhat underestimates the planetary flux. On the nightside, the same model overestimates the flux in both the 3.6 and the 4.5~$\mu$m bandpasses. Comparing the overall shape of the best-fit phase curves with that of the model-generated light curves (Figure~\ref{lightcurve1} and Table~\ref{tab:comparison1}), we find that the calculated maximum and minimum flux offsets are overestimated by the $1\times$ solar no TiO/VO model; in other words, the SPARC model predicts a larger eastward shift in the location of the hot and cold regions in the planet's dayside and nightside atmosphere, respectively. The phase curve amplitude at 4.5~$\mu$m is well matched by the model, while the amplitude at 3.6~$\mu$m is overestimated.

One explanation for the lower-than-predicted nightside planetary emission on WASP-19b is the presence of high-altitude clouds in or above the photosphere.  A thick cloud would block the outgoing planetary flux from deeper, warmer levels, thereby suppressing the measured planetary emission at all wavelengths. In this case, the detected emission would resemble blackbody radiation from the cold cloud tops. Indeed, as discussed in Section~\ref{subsec:temps}, we find that the measured nightside spectrum for WASP-19b is well described by a single blackbody with an effective temperature of $1063^{+195}_{-250}$~K. The presence of clouds also means that observations of the nightside atmosphere would primarily probe higher altitudes in the atmosphere than in the cloud-free case. At these lower pressure levels, the radiative timescales are shorter, corresponding to less efficient day--night recirculation and a smaller predicted eastward offset in the location of minimum flux. Since our sinusoidal phase curve model forces the maximum and minimum fluxes to have the same phase offset ($c_{2}$), this scenario may cause the overall phase curve solution to predict smaller flux offsets than in the cloud-free case.

Another possible mechanism for producing lower nightside planetary fluxes is a higher atmospheric metallicity. The original SPARC models of \citet{showman2009} explored $5\times$ and $10\times$ solar atmospheric metallicities for HD 189733b and showed that high-metallicity implies larger day--night temperature differences compared to the $1\times$ solar case due to the higher optical opacity of high-metallicity atmospheres \citep[e.g.,][]{fortneymarley}. \citet{lewis2010} ran atmospheric circulation models for the hot Neptune GJ 436b and also found that high-metallicity models exhibit stronger day--night temperature contrasts than low-metallicity models. High metallicity models were found to provide a better match to the near-infrared spectroscopic phase curves of WASP-43b \citep{kataria}. 

To explore the effects of high atmospheric metallicity on the planetary emission of WASP-19b, we compare our observations to a circulation model assuming $5\times$ solar elemental ratios of heavy elements. As shown in Figures~\ref{lightcurve1} and \ref{jonathan1}, the high-metallicity model provides a very good match with the measured dayside planetary emission spectrum across all bandpasses. Meanwhile, the same model predicts lower nighttime planetary fluxes than the no TiO/VO model assuming solar metallicity: the high-metallicity model matches the measured nightside emission at 4.5~$\mu$m, while still overestimating the emission at 3.6~$\mu$m, albeit less severely than the case of solar metallicity. Comparing the shapes of the model light curves with those of our best-fit phase curve solutions, we find that the high-metallicity SPARC model predicts flux maxima, phase curve amplitudes, and flux offsets that are consistent with the corresponding values from the data in both bandpasses (at better than the $2\sigma$ level in all cases; see Table~\ref{tab:comparison1}). Overall, the model with no TiO/VO and $5\times$ solar atmospheric metallicity yields a better match to the data than the no TiO/VO model with solar metallicity.

For HAT-P-7b, the calculated dayside emission across all infrared bands is well-matched by the predicted spectrum from the SPARC model with TiO/VO, which suggests the presence of a dayside temperature inversion (Figure~\ref{jonathan2}). As shown in Figure~\ref{lightcurve2} and Table~\ref{tab:comparison2}, our best-fit maximum and minimum flux offsets are consistent with zero, while both the models with and without TiO/VO yield significant eastward shifts on the hotspot relative to the substellar point. The with TiO/VO model yields a less severe overestimate of the maximum and minimum flux offsets, with an overall phase curve shape that is more consistent with the \textit{Spitzer} data than the no TiO/VO model. 

The measured HAT-P-7b nightside emission is highly discrepant from the predictions of both SPARC models. At 3.6~$\mu$m, the observed nightside flux is very low, consistent with zero, while the 4.5~$\mu$m nightside planetary flux is higher than either of the models. This behavior is identical to that reported in our analysis of the 3.6 and 4.5~$\mu$m phase curves of the highly irradiated hot Jupiter WASP-14b \citep{wong2} as well as in previous observations of HAT-P-2b \citep{lewis}. \citet{lewis2014} suggested that an enhanced C/O ratio might explain this discrepancy for HAT-P-2b, and in \citet{wong2} we made a similar argument for WASP-14b. Here, we invoke the same interpretation which was described in detail in \citet{wong2} and propose that a high atmospheric C/O ratio can explain both of these trends. Increasing the C/O ratio above solar values leads to a relative excess of CH$_{4}$ and a corresponding depletion of CO \citep{moses}. As a result, the atmospheric opacity is enhanced at 3.6~$\mu$m and simultaneously reduced at 4.5~$\mu$m. A higher-than-predicted atmospheric opacity results in a lower-than-predicted planetary brightness at the corresponding wavelengths (and vice versa). 

A high atmospheric C/O ratio may also lead to a relative depletion of water. However, unlike in the case of CO and CH$_{4}$, the absorption cross-section of water does not vary significantly across the 3.6 and 4.5~$\mu$m \textit{Spitzer} bandpasses. Therefore, a reduction in water would cause an overall reduction in the atmospheric opacity in both bands and cannot explain both the lower-than-predicted 3.6~$\mu$m planetary flux and the higher-than-predicted 4.5~$\mu$m planetary flux.

\subsubsection{One-dimensional model spectra}

\begin{figure}[t]
\begin{center}
\includegraphics[width=9cm]{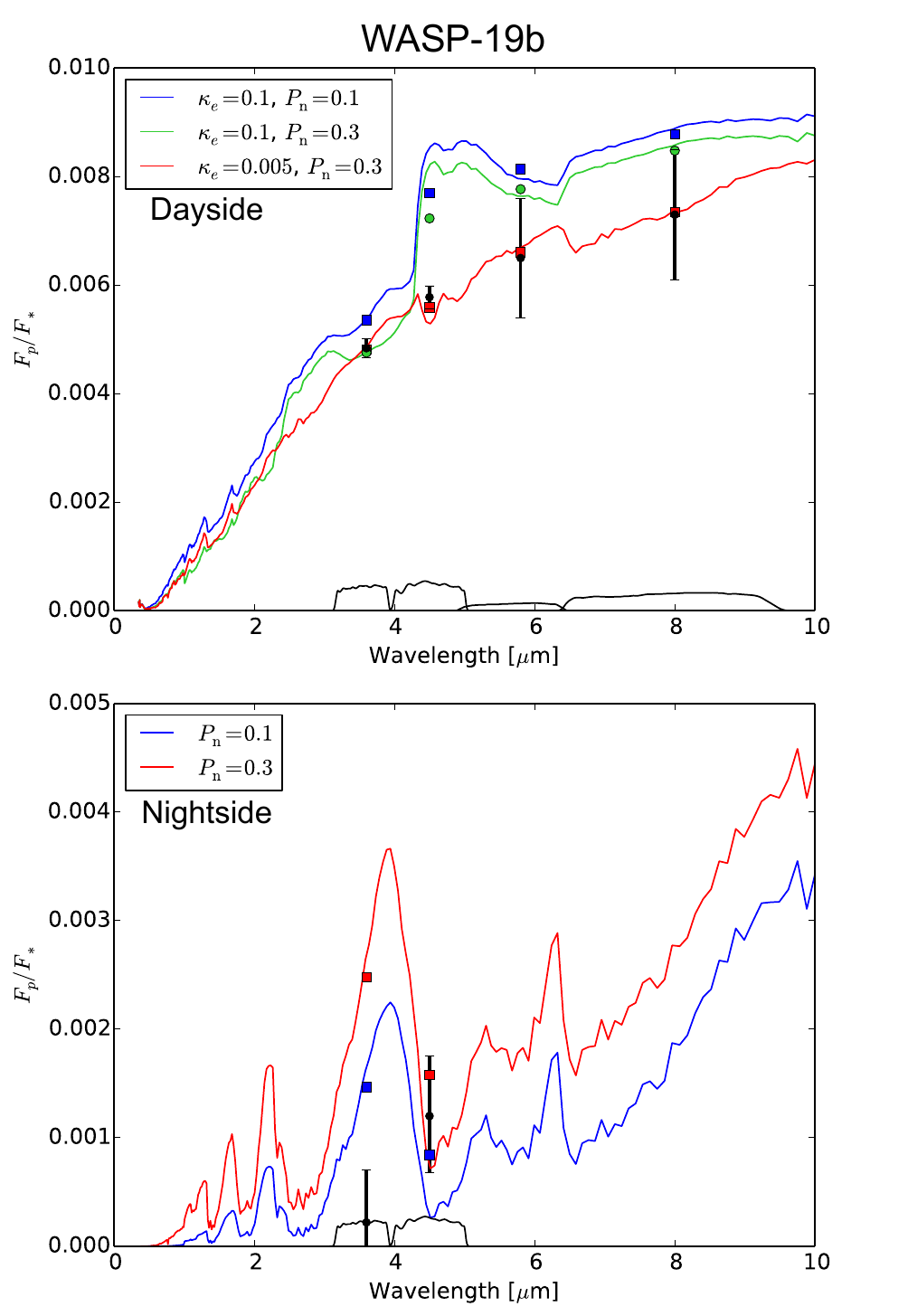}
\end{center}
\caption{Comparison of the error-weighted average dayside and nightside planet--star ratio (filled black circles) for WASP-19b with one-dimensional atmosphere model spectra following \citet{burrows}. The previouslypublished 5.8 and 8.0~$\mu$m eclipse depths reported in \citet{anderson} are also included. Solid colored lines indicate the model-generated spectra, with corresponding band-averaged points overplotted in the same color. The black lines at the bottom represent the photometric band transmission profiles in arbitrary units. The measured dayside planetary fluxes are consistent with the model with $\kappa_{e}=0.005$ and $P_{n}=0.3$, indicating that WASP-19b has no dayside thermal inversion and relatively efficient day--night recirculation. On the nightside, the measured nightside planetary fluxes in the 3.6 and 4.5~$\mu$m bands are not well-described by either the $P_{n}=0.1$ or the $P_{n}=0.3$ model. See text for description of model parameters.} \label{burrows1}
\end{figure}

\begin{figure}[t]
\begin{center}
\includegraphics[width=9cm]{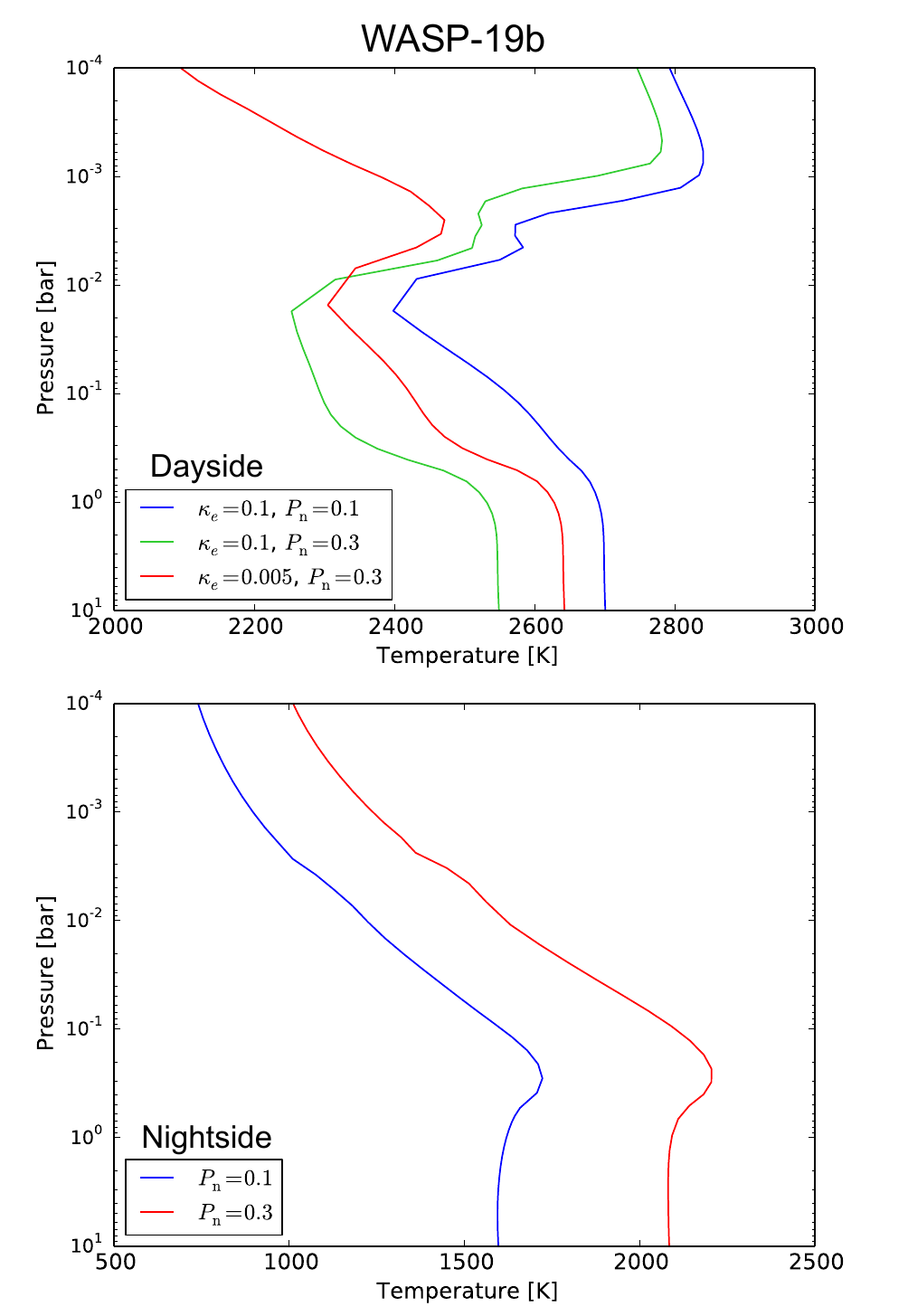}
\end{center}
\caption{Plot of temperature--pressure profiles for WASP-19b computed by the one-dimensional radiative transfer model used to  generate the dayside and nightside emission spectra in Figure~\ref{burrows1}.}\label{tp3}
\end{figure}

\begin{figure}[t]
\begin{center}
\includegraphics[width=9cm]{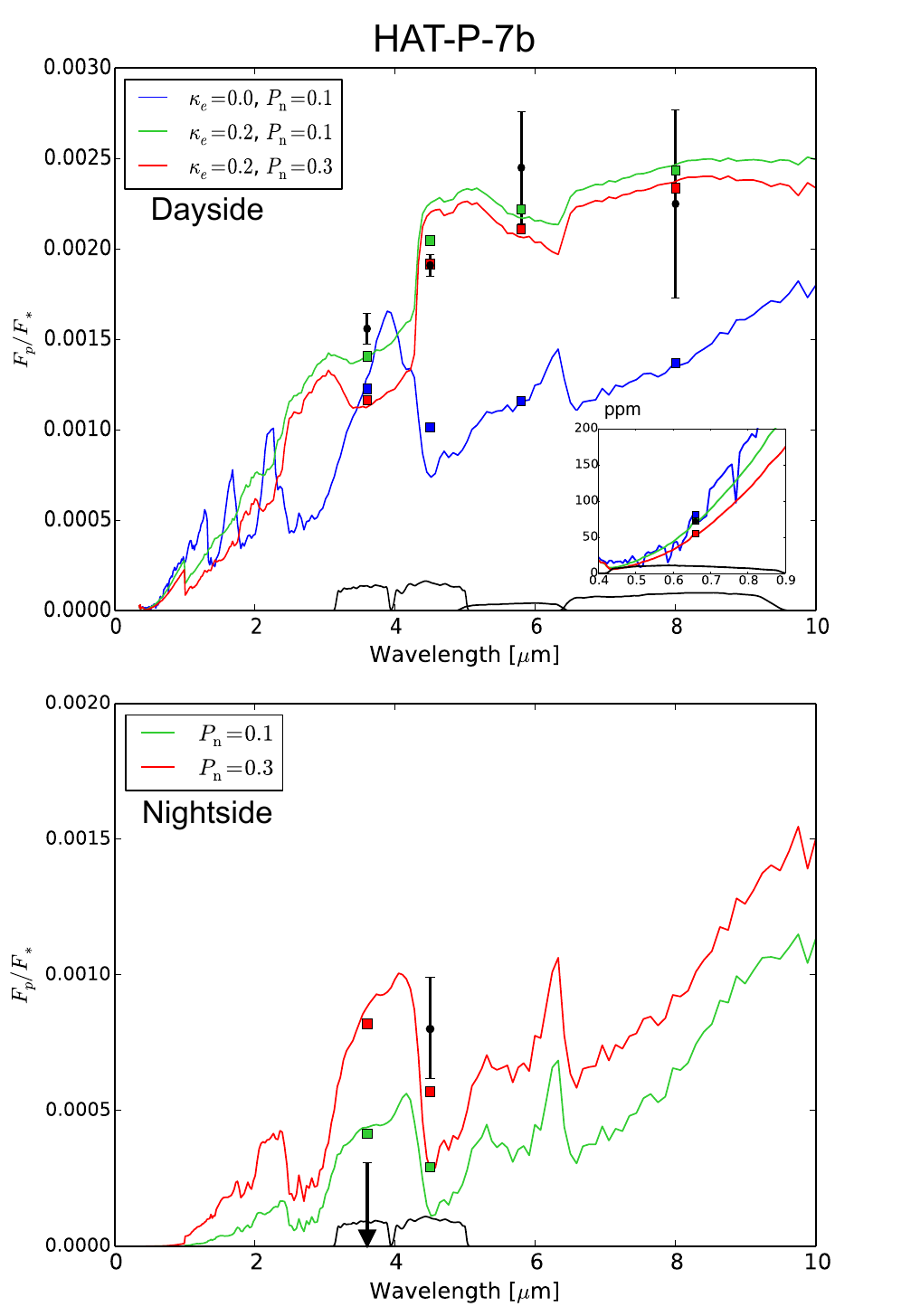}
\end{center}
\caption{Comparison of the broadband dayside and nightside planetary emission (filled black circles) for HAT-P-7b with one-dimensional atmosphere model spectra following \citet{burrows}. At 3.6~$\mu$m, the $2\sigma$ upper limit on the planetary brightness is shown.  The previously published 5.8 and 8.0~$\mu$m eclipse depths reported in \citet{christiansen} are also included. Solid colored lines indicate the model-generated spectra, with corresponding band-averaged points overplotted in the same color. The inset in the dayside comparison plot shows the measured eclipse depth in the \textit{Kepler} bandpass. The black lines at the bottom represent the photometric band transmission profiles in arbitrary units. The measured dayside planetary fluxes are most consistent with the model with $\kappa_{e}=0.2$ and $P_{n}=0.1$, indicating that HAT-P-7b has dayside thermal inversion and relatively inefficient day--night recirculation. On the nightside, the measured nightside planetary fluxes in the 3.6 and 4.5~$\mu$m bands are not well described by either the $P_{n}=0.1$ or the $P_{n}=0.3$ model. See text for a description of the model parameters.} \label{burrows2}
\end{figure}

For each planet, we run a suite of models following the methods of \citet{burrows} with a range of high-altitude optical absorber abundances (parametrized by $\kappa_{e}$, with units of cm$^{2}$~g$^{-1}$) and day--night energy recirculation efficiencies (parametrized by $P_{n}$, where $P_{n}=0.5$ indicates complete day--night redistribution and $P_{n}=0$ corresponds to re-redistribution on the dayside only.) Figure~\ref{burrows1} compares the model spectra for WASP-19b with the measured planetary emission; the corresponding temperature--pressure profile is shown in Figure~\ref{tp3}. We see that the $\kappa_{e}=0.005$ and $P_{n}=0.3$ model spectrum provides an excellent fit to the dayside data at all wavelengths. This indicates that WASP-19b has no thermal inversion in the dayside atmosphere and relatively efficient day--night heat transport, which is consistent with the conclusions from our comparison with 3D SPARC models. Meanwhile, on the nightside, the 3.6 and 4.5~$\mu$m flux ratios are not well-matched by either $P_{n}=0.1$ or the $P_{n}=0.3$ model. Here, we can once again invoke the presence of high silicate clouds on the nightside to explain the data, which are consistent with colder-than-predicted blackbody emission from the cloud tops.

\begin{figure}[t]
\begin{center}
\includegraphics[width=9cm]{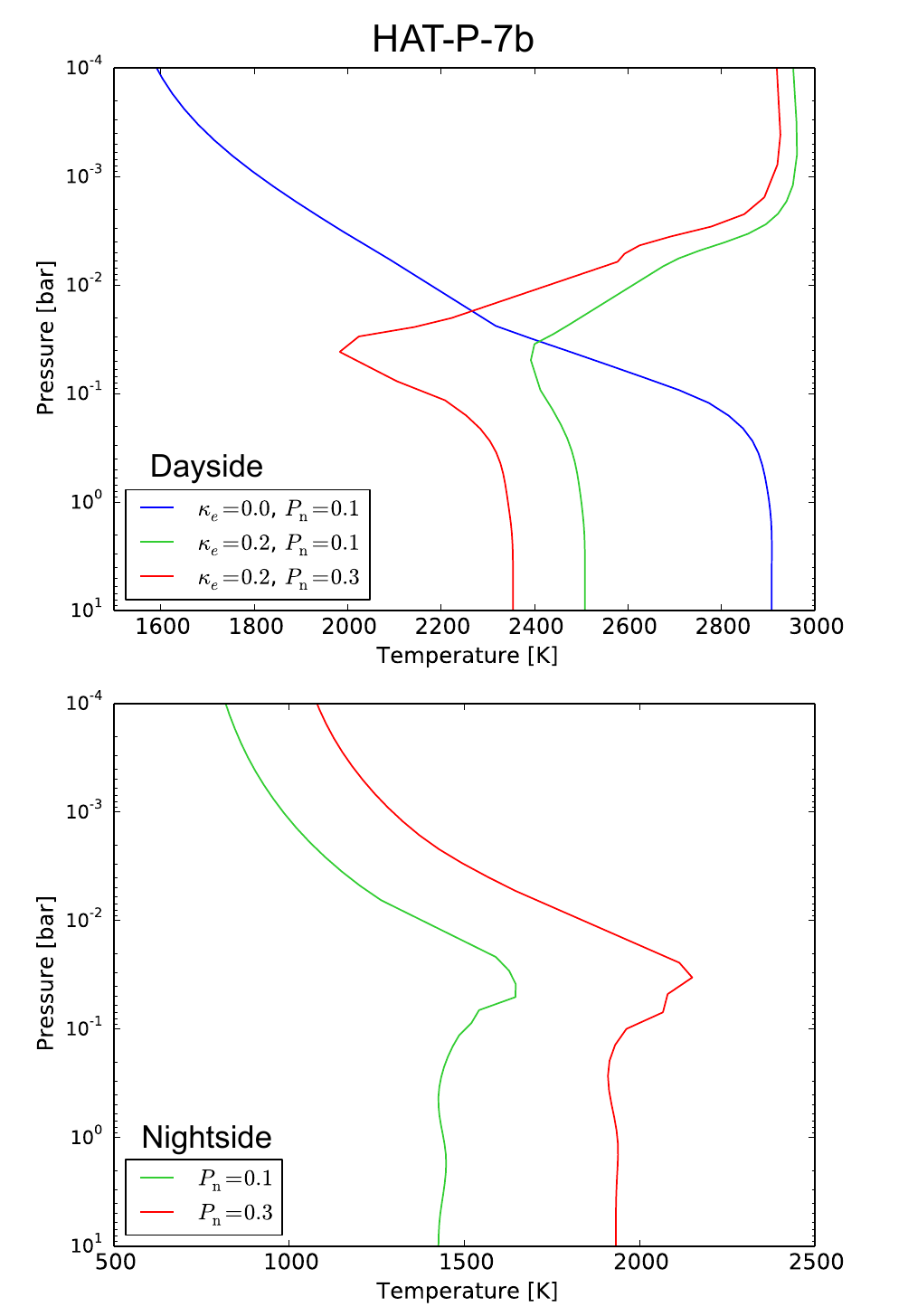}
\end{center}
\caption{Same as Figure~\ref{tp3}, for HAT-P-7b.}\label{tp4}
\end{figure}

The comparison between model spectra and planetary brightness data for HAT-P-7b is shown in Figure~\ref{burrows2}, with the corresponding temperature--pressure profile shown in Figure~\ref{tp4}. The $\kappa_{e}=0.2$ and $P_{n}=0.1$ model spectrum yields the best agreement with the measured dayside fluxes, which demonstrates that HAT-P-7b has a dayside temperature inversion and relatively poor day--night recirculation. These results match the conclusions from the SPARC model comparisons, as well as a previous comparison of the eclipse depths reported in \citet{christiansen} and 1D model spectra similar to the ones presented here \citep{spiegel}.  None of the 1D models reproduce both the very low measured 3.6~$\mu$m emission and the higher 4.5~$\mu$m flux ratio on the nightside. The nature of this discrepancy is identical to that of the mismatch between the SPARC model light curves and the observed nightside emission, as well as similar deviations between the 1D model spectra and the measured nightside planetary brightness reported for WASP-14b in \citet{wong2}, which provides support to the aforementioned hypothesis of an enhanced atmospheric C/O ratio on HAT-P-7b.

\subsection{Albedo}

\begin{figure}[t]
\begin{center}
\includegraphics[width=8cm]{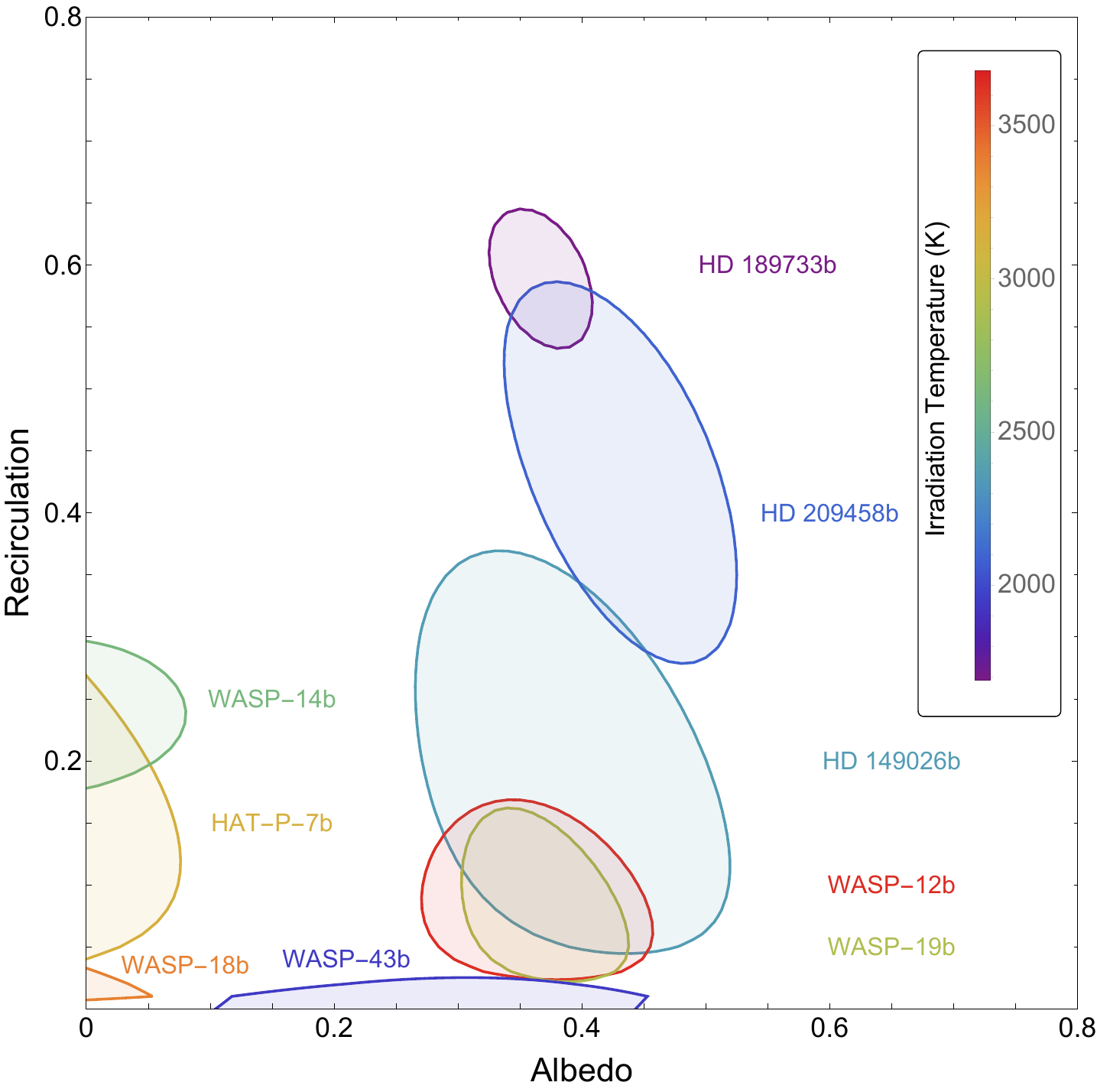}
\end{center}
\caption{Plot of $1\sigma$ Bond albedo and recirculation efficiency confidence regions for all planets with published thermal phase curve observations, following the methods of \citet{schwartz}.  The irradiation temperature is denoted by the color of the bounding curves. Both WASP-19b and HAT-P-7b have intermediate day--night heat transport, like other hot Jupiters of similar temperature. Two groups in albedo space are evident, with WASP-19b located in the higher-albedo group and HAT-P-7b lying in the lower-albedo group.} \label{planets}
\end{figure}

We  derive estimates for the albedo and recirculation using our best-fit eclipse depths and phase curve amplitudes. Following the methods described in \citet{schwartz}, we correct for contamination due to reflected starlight and solve for the Bond albedo $A_{B}$ and day--night heat transport efficiency $\epsilon$, where $\epsilon$ is defined such that $\epsilon=0$ indicates no heat recirculation to the nightside and $\epsilon=1$ signifies complete redistribution. For WASP-19b, we find $A_{B}=0.38\pm 0.06$ and $\epsilon=0.09\pm0.06$, while for HAT-P-7b, we obtain $A_{B}=0$ ($<0.08$ at $1\sigma$) and $\epsilon = 0.15\pm0.11$. Figure~\ref{planets} shows the location of WASP-19b and HAT-P-7b in albedo-recirculation space along with the seven other exoplanets with measured thermal phase curves. The relatively low day--night heat transport efficiencies and relatively high irradiation temperatures of WASP-19b and HAT-P-7b fit within the general observed trend that the efficiency of day--night recirculation is anti-correlated with the level of irradiation \citep{cowanagol2011,perez-becker}.

From Figure~\ref{planets}, two groupings in albedo space are apparent. This bifurcation was first reported in \citet{wong2}, at which point the low-albedo group was populated by WASP-14b and WASP-18b. Those planets are both significantly more massive ($7-10$ vs. $\sim 1~M_{\mathrm{Jup}}$) and have higher surface gravity than the other planets in the sample. The low albedo of high-gravity planets may be due to less clouds or hazes at the pressures probed by the $Spitzer$ bandpasses: the Stokes settling velocity is proportional to $g$, while the scale height is inversely proportional to $g$, so that the time required for a particle of a given size to gravitationally settle over a scale height scales as $g^{-2}$. In addition, both WASP-14b and WASP-18b are relatively young, and so we suggested in \citet{wong2} that this additional internal flux may be due to significant residual heat of formation.

With the inclusion of HAT-P-7b in the low-albedo group, these simple interpretations are no longer tenable. HAT-P-7b has a mass of $1.671\pm0.026~M_{\mathrm{Jup}}$, which is comparable with the masses of planets in the higher-albedo group. Moreover, the age of the HAT-P-7 system is estimated at $2.07^{+0.28}_{-0.23}$~Gyr \citep{lund}, making it older than most of the planets in the plotted sample. Although HAT-P-7b is not as massive as WASP-14b or WASP-18b, it is still possible that the bifurcation in albedo is indicative of divergent thermal evolution histories for these three systems as compared to the rest of the hot Jupiter sample. We also note that there is some evidence from the \textit{Kepler} sample of hot Jupiters for a bimodal albedo distribution in the optical bandpass \citep{hengdemory}, as well as for spatially inhomogeneous clouds \citep{demory,hu,shporerhu}. The presence of clouds can suppress the infrared emission at the corresponding phases and affect our estimates of the albedo and recirculation from simple thermal energy budget considerations. We note that a recent modeling of HAT-P-7b's optical phase curve inferred a higher Bond albedo in the range $0.11 < A_{B} < 0.72$ \citep{vonparis}, consistent with the calculated albedo values for planets in the higher-albedo group. Ultimately, the albedos of hot Jupiters are likely the result of a complex interplay between characteristics of the planet and/or the host star, atmospheric chemistry and dynamics, as well as different formation/migration histories.

\section{Conclusions}
In this paper, we present an analysis of full-orbit 3.6 and 4.5~$\mu$m phase curve observations of the highly irradiated hot Jupiters WASP-19b and HAT-P-7b. For WASP-19b, we obtain error-weighted average best-fit secondary eclipse depths of  $0.485\%\pm 0.024\%$ and $0.584\%\pm 0.029\%$ at 3.6 and 4.5~$\mu$m. The dayside atmospheric brightness in the two bands is well fit by a single blackbody with an effective temperature of $2372^{+59}_{-60}$~K. For HAT-P-7b, we find that the best-fit 3.6 and 4.5~$\mu$m eclipse depths of $0.156\%\pm 0.009\%$ and $0.190\%\pm0.006\%$, respectively, are consistent with a single blackbody with an effective temperature of $2667\pm 57$~K. Our global phase curve fits and RV analyses produce improved estimates of the planets' orbital and physical parameters, including more precise estimates of the orbital periods: $P=0.788838989\pm0.000000040$~days (WASP-19b) and $P=2.2047372\pm0.0000011$ (HAT-P-7b).

We compare the results of our phase curve analysis with model spectra and light curves generated from both one-dimensional radiative transfer models and three-dimensional general circulation models (GCMs). The measured dayside planetary fluxes for WASP-19b suggest the absence of a temperature inversion as well as relatively efficient day--night heat transport; we detect a significant eastward shift in the hotspot relative to the substellar point, which is consistent with the predictions of the GCMs and indicative of a superrotating equatorial jet. The nightside planet--star flux ratios are not well-described by the solar thermochemical equilibrium composition models. Instead, they match the emission from a single blackbody with an effective temperature of $1090^{+190}_{-250}$~K. This may be indicative of the presence of high-altitude silicate clouds in the nightside atmosphere and/or an enhanced atmospheric metallicity. Pursuing the latter idea, we show that a GCM with $5\times$ solar metallicity provides a better agreement with the measured dayside emission spectrum and overall phase curve shape than the case of solar metallicity.

In contrast with WASP-19b, the dayside planetary emission for HAT-P-7b derived from \textit{Spitzer} data is consistent with a temperature inversion in the dayside atmosphere and relatively inefficient day--night recirculation. From the calculated maximum and minimum flux offsets, we do not detect any eastward shift in the hotspot from the substellar point. Meanwhile, the measured nightside planetary emission at 3.6 and 4.5~$\mu$m differs from both the one-dimensional and the three-dimensional models in a manner similar to that reported for WASP-14b in \citet{wong2}. Specifically, the very low 3.6~$\mu$m nightside planetary flux indicates a significantly higher atmospheric opacity at that wavelength than is predicted by the models and suggests an enhanced C/O ratio.

While it is not possible to measure the C/O ratio of hot Jupiter atmospheres directly with current telescope facilities, one can  instead obtain measurements of the host star's C/O ratio as a proxy. HAT-P-7 has an intermediate measured C/O ratio of $0.42 \pm 0.14$ \citep{teske}, which is consistent with solar. Nevertheless, the detailed formation and migration history of hot Jupiters in general can result in a wide range of C/O ratios, which is primarily determined by the region of the gas disk from which most of the atmosphere is accreted \citep[e.g.,][]{oberg,alidib,helling,madhusudhan}. Further observations of HAT-P-7b at different wavelengths, possibly with the {\it Hubble Space Telescope}'s Wide Field Camera 3, may offer important additional constraints on the atmospheric composition. In the context of theoretical and numerical studies, the question of non-solar C/O ratios and how they affect the global circulation patterns has not yet been addressed in current three-dimensional GCMs and merits further exploration. More broadly, the burgeoning body of phase curve studies has revealed a wide diversity of observed behaviors, many of which diverge from the predictions of current models and may require a more in-depth consideration of phenomena such as clouds and disequilibrium chemistry to adequately explain.

We use a simple thermal balance model to place WASP-19b and HAT-P-7b in the context of other planets with full-orbit thermal measurements and find that both of these highly irradiated atmospheres have relatively low recirculation efficiencies when compared with cooler planets. The emerging bimodality seen in the distribution of Bond albedos remains an open question; both WASP-19b and HAT-P-7b have similar masses and irradiation temperatures, yet their albedos are notably discrepant, refuting the correlation between low albedo and high planet mass posited in \citet{wong2}. Further study into the formation and evolution models for hot Jupiter atmospheres of various chemical compositions promises to expand our knowledge of the relevant planetary and/or stellar properties that determine a planet's albedo.

\qquad

This work is based on observations made with the Spitzer Space Telescope, which is operated by the Jet Propulsion Laboratory, California Institute of Technology under a contract with NASA. Support for this work was provided by NASA through an award issued by JPL/Caltech. This work is also based in part on observations obtained at the W.M. Keck Observatory using time granted by the University of Hawaii, the University of California, and the California Institute of Technology.  We thank the observers who contributed to the measurements reported here and acknowledge the efforts of the Keck Observatory staff.  We extend special thanks to those of Hawaiian ancestry on whose sacred mountain of Mauna Kea we are privileged to be guests. This work was performed in part under contract with the Jet Propulsion Laboratory (JPL) funded by NASA through the Sagan Fellowship Program executed by the NASA Exoplanet Science Institute.

\end{document}